\documentclass[
prd,aps,amsmath,superscriptaddress,nofootinbib,
]{revtex4}

\usepackage[hypertex]{hyperref}
\usepackage{graphicx}
\usepackage{epstopdf}
\usepackage{bm}
\usepackage{epsfig}
\usepackage{xspace}
\usepackage{color}

  \newcount\hour \newcount\minute
  \hour=\time \divide \hour by 60
  \minute=\time
  \newcommand{\mydate}{\ \today \ - \number\hour :\ifnum \minute<10 0\fi 
\number\minute}

\def\OMIT#1{}

\newcommand{\nn}{\nonumber} 

\newcommand{\bn}{{\bar n}}
\newcommand{\bea}{\begin{eqnarray}}
\newcommand{\eea}{\end{eqnarray}}

\newcommand{\mcdot}{\!\cdot\!}

\newcommand{\plus}{\ensuremath{\! + \!}}
\newcommand{\minus}{\ensuremath{\! - \!}}

\def\lsim{\mathrel{\raise.3ex\hbox{$<$\kern-.75em\lower1ex\hbox{$\sim$}}}}
\def\gsim{\mathrel{\raise.3ex\hbox{$>$\kern-.75em\lower1ex\hbox{$\sim$}}}}

\begin{document}
\setlength\baselineskip{12pt}



\title{Hemisphere Soft Function at ${\cal{O}}(\alpha_s^2)$ 
for Dijet Production \\[2mm]
in $e^+e^-$ Annihilation \\[3mm]\mbox{}} 

\vspace*{1cm}

\author{Andre H.~Hoang and Stefan Kluth}
  \affiliation{Max-Planck-Institut f\"ur Physik\\ (Werner-Heisenberg-Institut)\\
  F\"ohringer Ring 6,  80805 M\"unchen, Germany, \footnote{Electronic address: 
 ahoang@mppmu.mpg.de, skluth@mppmu.mpg.de}\\[10mm]\mbox{}}




\begin{abstract}
\vspace{0.4cm}
\setlength\baselineskip{12pt}
We determine the ${\cal{O}}(\alpha_s^2)$ corrections to the partonic
hemisphere soft function relevant for thrust and jet mass distributions in
$e^+e^-$ annihilation in the dijet limit. In this limit the distributions can be
described by a factorization theorem that sums large logarithmic terms and
separates perturbative from nonperturbative effects. Using the known
${\cal{O}}(\alpha_s^2)$ contributions of the jet functions and the hard 
coefficients in the factorization theorem, constraints from renormalization
group evolution and nonabelian exponentiation, and results from
numerical integration of ${\cal{O}}(\alpha_s^2)$ QCD matrix elements,
the ${\cal{O}}(\alpha_s^2)$ corrections of the soft function can be
determined unambiguously. We study  
the impact of subtracting contributions related to the 
${\cal{O}}(\Lambda_{\rm QCD })$ renormalon in the partonic threshold using the
soft function gap proposed recently by Hoang and Stewart, and we discuss the
importance to account for the renormalization group evolution of
the gap parameter. As a byproduct we also present the previously unknown 
next-to-next-to-leading logarithmic anomalous dimensions for the hard
coefficient that appear in the factorization theorem for the
double differential invariant mass distribution for heavy quark pair
production at high energies in the resonance region proven by Fleming etal.
\\[25pt]
 \hbox{MPP-2008-25}\\
 \hbox{arXiv:0806.3852 [hep-ph]}
\end{abstract}

\maketitle

\newpage

\newpage
\section{Introduction}
\label{sectionintroduction}

One of the most prominent and important features of hadronic final states
produced in particle collisions is the jet structure, i.e.\,the presence of a
small number of collimated groups of particles recoiling against each
other. Quantifying the structure of hadronic final states in $e^+e^-$
annihilation in terms of event shapes allows for a direct comparison of
experimental data with predictions in QCD and thus for precise measurements
of parameters and stringent test of the theory~\cite{Kluth:2006bw}.  
Event shape variables avoid direct association of particles to individual jets
and calculate instead a single number that accounts for, and classifies the
event according to its jet topology. Generally, the observables are
constructed such that a value close to zero corresponds to a dijet event
topology where two jets of energetic particles are produced back-to-back with additional
soft particles between the jets. Most events enter in this dijet
region of the event shape distributions because two
jets can already be produced at tree-level while three and more jets are
suppressed by additional powers of the strong coupling. 

Among the most common event shape variables are the thrust $T$ defined
by~\cite{Farhi:1977sg} 
\begin{align}
\label{Tdef}
T \,\equiv \,
{\rm max}_{\hat n}\left(
\frac{\sum_i |\vec p_i\cdot \hat n| }{ \sum_i |\vec p_i| }
\right)\,,
\end{align}
where the sum is over all particles, and $\vec p_i$ is the three-momentum of
particle $i$. The thrust axis is the unit vector $\hat n$ which maximizes the
expression in the parentheses. Since $T\sim 1$ characterizes  dijet-events it
is convenient to use the variable
\begin{equation}
\label{taudef1}
\tau \,\equiv\, 1-T
\end{equation}
as the thrust variable. The plane through the interaction point and
perpendicular to the thrust axis divides the event into two hemispheres, $H_1$
and $H_2$. The maximum of the squared invariant masses $M_1^2$ and $M_2^2$
of all the particles in $H_1$ and $H_2$, respectively, defines the heavy jet mass
variable~\cite{Chandramohan:1980ry}
\begin{align}
\label{rhodef}
\rho\,\equiv\,
\frac{{\rm max}(M_1^2,M_2^2)}{Q^2}\,,
\end{align}
where $Q$ is the $e^+e^-$ c.m.\,\,energy. In the dijet limit where $\tau\ll 1$,
the thrust can be written as a function of the hemisphere Masses $M_1$ and
$M_2$,
\begin{align}
\label{taudef2}
\tau \, = \, \frac{M_1^2+M_2^2}{Q^2}\, + \, 
{\cal O}\left(\frac{M_{1,2}^4}{Q^4}\right)\,.
\end{align}
In fact, the double-differential $M_1$-$M_2$ invariant mass distribution 
represents by itself an event shape distribution, where small values
of $M_{1,2}^2/Q^2$ correspond to the dijet region. 

For massless quarks in the dijet region the $M_1$-$M_2$
distribution can be described by a factorization
theorem~\cite{Korchemsky:1998ev,Korchemsky:1999kt,Bauer:2003di,
Fleming:2007qr,Fleming:2007xt,Bauer:2008dt} 
\begin{align}
\label{facttheoSCET}
\frac{d^2 \sigma^{\rm dijet}}{dM^2_1 dM^2_2} &= 
\sigma_0 \, H_Q(Q\OMIT{,\mu_Q},\mu)
\,\, \int^\infty_{-\infty}\!\!\! d\ell^+  d\ell^- 
J_1(M_1^2-Q\,\ell^+,\mu)\,
J_2(M_2^2-Q\,\ell^-,\mu)\,
S(\ell^+, \ell^-,\mu) \,,
\end{align}
which is valid at leading order in $M_{1,2}^2/Q^2$. Here, $\sigma_0$ is the
tree level total cross section, $H_Q$ is a calculable hard coefficient and
$J_{1,2}$ are calculable jet functions, whereas $S(\ell^+,\ell^-,\mu)$ is the
hemisphere soft function. A similar factorization theorem for the
$M_1$-$M_2$ distribution can be derived for the production of quarks with mass
$m\gg\Lambda_{\rm QCD}$. Here the dijet region corresponds to the region
near the heavy quark mass resonance, where 
$\hat s_{1,2}\equiv (M_{1,2}^2-m^2)/m\ll m$. The corresponding
factorization theorem has the 
form~\cite{Fleming:2007qr,Fleming:2007xt}    
\begin{align} 
\label{facttheobHQET}
  \frac{d^2 \sigma^{\rm dijet}}{ dM_1^2\, dM_2^2} &= 
  \sigma_0 \: H_Q(Q,\mu_m)\,H_m\Big(m,\frac{Q}{m},\mu_m,\mu\Big)
  \int\! d\ell^+ d\ell^- B_+\Big(\hat s_1-\frac{Q\ell^+}{m},\mu\Big)\:
  B_-\Big(\hat s_2-\frac{Q\ell^-}{m},\mu\Big) 
  S(\ell^+\!,\ell^-\! ,\mu) \,,
\end{align}
which is valid at leading order in $m^2/Q^2$ and
$(\hat s_{1,2}+\Gamma_q)/m$.  Here 
$H_m$ is a calculable hard coefficient and $B_\pm$ are calculable jet
functions for heavy quarks describing invariant mass fluctuations below the
scale $m$. The term $\Gamma_q$ is the heavy quark width.
The soft function $S$ is equivalent to the one in Eq.~(\ref{facttheoSCET}).
As a consequence of the separation of the different physical modes in the
factorization theorems~(\ref{facttheoSCET}) and 
(\ref{facttheobHQET}), the hard coefficients, and the jet and soft functions
are renormalization scale dependent. 

The soft function carries information on how the soft radiation between the
two energetic jets is associated to the invariant masses $M_1$ and $M_2$. For
the hemisphere prescription described above it is defined
as~\cite{Korchemsky:1998ev,Korchemsky:1999kt,Korchemsky:2000kp,Fleming:2007qr,Bauer:2008dt} 
\begin{align} \label{Sdef1}
  S(\ell^+,\ell^-,\mu) &\equiv \frac{1}{N_c}\sum _{X_s} \delta(\ell^+\! \minus
  k_s^{+a}) \delta(\ell^- \!\minus k_s^{-b}) \langle 0| (\overline
  {Y}_\bn)^{cd}\, ({Y}_n)^{ce} (0) |X_s \rangle \langle X_s|
  ({Y}^\dagger_n)^{ef}\, (\overline {Y}_\bn^\dagger)^{df} (0) |0\rangle .
\end{align}
Here $k_s^{+a}$ is the total plus-momentum of soft hadrons in $X_s$ that are
in hemisphere 1, $k_s^{-b}$ is the total minus momentum for soft hadrons in
the other hemisphere. The soft function for thrust is related to the
hemisphere soft function by 
\begin{align}
\label{Sthrust}
S_T(\tau,\mu)=\int d\ell^+
d\ell^-\delta\Big(\tau-\frac{\ell^++\ell^-}{Q}\Big)\,S(\ell^+,\ell^-,\mu)\,.
\end{align}
The definition of the soft function only depends on the light-like kinematics
and the color state of the primary quark-antiquark pair and thus can be
written in terms of the Wilson lines
\begin{align} 
  Y_n^\dagger(x) &=   {\rm P} \, 
   \exp\Big(i g\! \int_{0}^\infty \!\!\!\!ds\, n\mcdot A_{s}(ns\!+\!x) \Big) \,,
 & \overline {Y_\bn}^\dagger(x) & =   {\rm P} \: \exp\Big( i g\! \int_{0}^{\infty} \!\!\!\!ds\, 
      \bn\mcdot \overline {A}_{s}(\bn s\!+\! x) \Big) 
  \,.
\end{align}
In general, $S(\ell^+,\ell^-,\mu)$ is a nonperturbative function that peaks for
$\ell^\pm\sim\Lambda_{\rm QCD}$ when $\mu\gsim\Lambda_{\rm QCD}$. Depending on
the size of $M_{1,2}$ different aspects of the soft function are important
since the convolutions in Eqs.~(\ref{facttheoSCET}) and (\ref{facttheobHQET})
probe momenta $\ell^\pm\sim M_{1,2}^2/Q$ and $\sim \hat s_{1,2}m/Q$,
respectively. In the immediate resonance region we have $M_{1,2}^2\sim
Q\Lambda_{\rm QCD}$ and $\hat s_{1,2}\sim Q\Lambda_{\rm
QCD}/m + \Gamma_q$~\cite{Fleming:2007qr}\footnote{The heavy quark width
$\Gamma_Q$ plays a vital role for top quarks, but can be neglected for bottom
quarks.},  
and the nonperturbative distribution described by $S(\ell^+,\ell^-,\mu)$ affects
directly the shape of the differential cross section. Here, the soft function
can be written as a convolution of the partonic soft function, computed in
perturbation theory at a scale $\mu=\mu_\Delta\gsim\Lambda_{\rm QCD}$, with a
nonperturbative model function that can be determined from experimental
data~\cite{Hoang:2007vb}, 
\begin{align}
\label{Sdef2}
S(\ell^+,\ell^-,\mu) & = 
\int_{-\infty}^{+\infty}\!\!\! d\ell^{\prime +}
\int_{-\infty}^{+\infty}\!\!\! d\ell^{\prime -}\
S_{\rm part}(\ell^+ \minus \ell^{\prime +},\ell^- \minus \ell^{\prime -},\mu)\,
S_{\rm mod}(\ell^{\prime +},\ell^{\prime -}) 
\,.
\end{align}
In the tail region away from the resonance we have $M_{1,2}^2\gg
Q\Lambda_{\rm QCD}$ and $\hat s_{1,2}\gg Q\Lambda_{\rm QCD}/m+\Gamma_q$ and the
soft function can be determined from the partonic soft function plus power
corrections that can be also related to Eq.~(\ref{Sdef2}). Note that the
partonic soft function contains $\delta$ functions and plus-distributions of
the variables $\ell^\pm$, so that the lower limit of the
$\ell^\pm$-integrations in 
Eq.~(\ref{Sdef2}) is zero. In Ref.~\cite{Korchemsky:2000kp} the function
\begin{align} \label{fmod1}
  f_{\rm exp}(\ell^{\prime +},\ell^{\prime -}) \, = \,
   \theta(\ell^{\prime +})\,\theta(\ell^{\prime -})\,
   \frac{ {\cal N}(a,b) }{\Lambda^2}
  \Big( \frac{\ell^{\prime +}\ell^{\prime -}}{\Lambda^2}\Big)^{a-1} \exp\Big(
  \frac{-(\ell^{\prime +})^2-(\ell^{\prime -})^2-2 b
    \ell^{\prime +}\ell^{\prime -}}{\Lambda^2} \Big)
\,,
\end{align}
was suggested as a two-parameter model for $S_{\rm mod}$. 
Here ${\cal N}(a,b)$ is a factor that is chosen such that the integral of
$f_{\rm exp}$ of the positive $\ell^{\prime \pm}$ plane is unity. This
normalization is required by the consistency of Eq.~(\ref{Sdef2}) with the
power expansion in the tail region. The ${\cal
O}(\alpha_s)$ corrections to the partonic hemisphere soft function were
computed in Ref.~\cite{Fleming:2007xt} (see also Ref.~\cite{Schwartz:2007ib}
for the ${\cal O}(\alpha_s)$ corrections to the partonic soft function for the
thrust distribution).  

It has been noticed in
Refs.~\cite{Gardi:2001ny,Hoang:2007vb} that the partonic threshold of the
soft function
$S_{\rm part}(\ell^\pm,\mu)$ at $\ell^\pm=0$ has an ${\cal O}(\Lambda_{\rm
QCD})$ renormalon ambiguity that is similar in nature to the well-known ${\cal
O}(\Lambda_{\rm QCD})$ pole mass renormalon. In Ref.~\cite{Hoang:2007vb} is
was shown that this ambiguity can be removed by introducing a gap in the soft
function model such that it vanishes for $\ell^{\prime \pm} < \Delta$. A way
to achieve this is by defining
\begin{align} \label{Sgap}
  S_{\rm mod}(\ell^{\prime +},\ell^{\prime -})
  \,  \equiv \,  
  f_{\rm exp}(\ell^{\prime +}-\Delta,\ell^{\prime -} - \Delta) \,.
\end{align}
Here $\Delta$ can be interpreted as the minimum hadronic energy deposit in
each hemisphere. Via the convolution in Eq.~(\ref{Sdef2}) the term $\Delta$
compensates the renormalon ambiguity in $S_{\rm part}$ at its partonic
threshold. It is then possible to explicitly remove this renormalon by writing
$\Delta = \bar\Delta+\delta\bar\Delta$, where $\bar\Delta$ is free of an
${\cal O}(\Lambda_{\rm QCD})$ renormalon
and $\delta\bar\Delta$ is a perturbation series that cancels the renormalon
ambiguity in the partonic soft function
order-by-order. It was demonstrated in Ref.~\cite{Hoang:2007vb} that this
procedure leads to a substantial reduction of the perturbative uncertainties
that come from the soft function. 

In this paper we determine the ${\cal O}(\alpha_s^2)$ two-loop corrections to
the partonic hemisphere soft function $S_{\rm part}$, which previously was the
only term in the factorization theorems~(\ref{facttheoSCET}) and
(\ref{facttheobHQET}) that was not known at ${\cal O}(\alpha_s^2)$. The result
enables a full next-to-next-to-leading logarithmic (NNLL) determination of the 
distributions described by the factorization theorems~(\ref{facttheoSCET}) and
(\ref{facttheobHQET}).  
Using constraints on the form of the soft function from its renormalization
group (RG) properties, the nonabelian exponentiation
theorem~\cite{Gatheral:1983cz,Frenkel:1984pz} and from the 
$1\leftrightarrow 2$ symmetry with respect to the hemispheres $H_1$ and $H_2$,
it is possible to determine the analytic form of the previously unknown ${\cal
O}(\alpha_s^2)$ contributions of the soft function up to two constants.
These constants can be determined from Eq.~(\ref{facttheoSCET}) by taking the
known two-loop results for the hard coefficient
$H_Q$~\cite{Matsuura:1988sm,Moch:2005id,Gehrmann:2005pd} 
and  the jet function $J_{1,2}$ ~\cite{Becher:2006qw} as an input and
using numerical results from the MC program
EVENT2~\cite{Catani:1996jh,Catani:1996vz} for thrust and heavy 
jet mass distributions for massless quarks at ${\cal O}(\alpha_s^2)$.
The result we obtain are confirmed by correct predictions for
distributions of variants of the thrust and heavy jet mass variables that can
also be obtained from EVENT2. To our knowledge  these variants of the thrust
and heavy jet mass variables have not been defined in the literature before. 

The program of this paper is as follows. In Sec.~\ref{sectionproperties} we
review the previously known properties of the partonic soft function. We show
that these constraints determine the two-loop partonic soft function 
up to two unknown constants that can be determined numerically. In
Sec.~\ref{sectiondistribution} we determine the ${\cal O}(\alpha_s^2)$
expressions for the different event shape distributions employed in our
numerical work and in Sec.~\ref{sectionevent2} we present the analysis to
determine the two constants from EVENT2. In Sec.~\ref{sectiongap} we introduce
a renormalon-free gap parameter $\bar\Delta$ from the position space soft
function. This gap parameter depends on an infrared scale $R$ and the
renormalization scale $\mu$, and we determine its evolution equations in $R$
and $\mu$. Section~\ref{sectionnumerical} contains a 
brief numerical analysis illustrating the impact of using the renormalon-free
gap parameter and of accounting properly for its scale dependences. The
conclusions are presented in 
Sec.~\ref{sectionconclusions}. We have attached three appendices collecting
useful formulae on the Fourier transform of plus-distributions, on results for
the hard coefficient $H_Q$ and the massless jet function $J$ adapted to our
notation, and on the cumulative event shape distributions that are used in
this work. In particular, App.~\ref{app:hardjetcc} contains a determination of
the previously unknown non-cusp NNLL anomalous dimension of the hard
coefficient $H_m$ that appears in the factorization theorem for massive jets
in Eq~(\ref{facttheobHQET}).

\section{Properties of the Hemisphere Soft Function}
\label{sectionproperties}

In this section we summarize the known properties of the hemisphere soft
function. These properties lead us to a particular analytic form for the
previously unknown parts of the partonic soft function at ${\cal
  O}(\alpha_s^2)$ that depend only 
on two unknown parameters. These parameters are determined numerically in
Sec.~\ref{sectionevent2}. To simplify the notation and avoid cluttering due to
convolution integrals in the variables $\ell^\pm$, we use in this work mainly
the position space representation. A number of formulas in the
$\ell^\pm$ momentum space variables useful for future applications
are collected in the appendix. In
position $x_{1,2}$-space the soft hemisphere function is defined as,  
\begin{align}
\label{Sxdef1}
S(x_1,x_2,\mu) \, = \, S(x_2,x_1,\mu) \, = \,
\int\!\! d\ell^{+} d\ell^{-} \:
e^{-i \,\ell^+ x_1}\,e^{-i\,\ell^- x_2}\,
S(\ell^+,\ell^-,\mu)\,.
\end{align}
The soft function is $x_1\leftrightarrow x_2$ symmetric because the definition
of the soft function in Eq.~(\ref{Sdef1}) is symmetric with respect to
exchanging the  hemispheres $H_1$ and $H_2$.  
\\

\noindent
{\it Result at ${\cal O}(\alpha_s)$.} It is straightforward to compute the
partonic hemisphere soft function at ${\cal O}(\alpha_s)$ from the definition
given in Eqs.~(\ref{Sdef1}). The result reads~\cite{Fleming:2007xt}
\begin{align}
\label{S1loop}
S_{\rm part}^{{\cal O}(\alpha_s)}(x_1,x_2,\mu)\, = \, 1 
- \frac{C_F\,\alpha_s(\mu)}{\pi}\left[\ln^2\left(i\,x_1
e^{\gamma_E}\mu\right)
+\frac{\pi^2}{8}\right]
- \frac{C_F\,\alpha_s(\mu)}{\pi}\left[\ln^2\left(i\,x_2
e^{\gamma_E}\mu\right)
+\frac{\pi^2}{8}\right]\,.
\end{align}

\noindent
{\it Renormalization Group Structure.} 
From the dijet factorization theorem for massless jets in
Eq.~(\ref{facttheoSCET}) one can derive 
consistency conditions~\cite{Fleming:2007xt} which relate the RG-evolution of
the soft function to the RG-evolution of the hard coefficient $H_Q$
and the jet function $J$. Details of the computation can be found in
App.~\ref{app:hardjetcc}. Given $S(x_1,x_2,\mu_0)$ at the scale $\mu=\mu_0$ we find
that~\cite{Fleming:2007xt}
\begin{align}
\label{UUS1}
S(x_1,x_2,\mu) \, = \, U_s(x_1,\mu,\mu_0)\,
U_s(x_2,\mu,\mu_0)\,S(x_1,x_2,\mu_0)\,, 
\end{align}
where 
\begin{align}
\label{Usdefx}
U_s(x,\mu,\mu_0)\, = \, 
\exp\bigg[\,
\tilde \omega(\Gamma_s,\mu,\mu_0)\,\ln\Big(  i\,x\,\mu_0 e^{\gamma_E} \Big)
 + \tilde K(\Gamma_s,\gamma_s,\mu,\mu_0)\,\bigg] \,,
\end{align}
with [\,$r\equiv\alpha_s(\mu)/\alpha_s(\mu_0)$\,]
\begin{align}
\label{omegaK}
\tilde \omega(\Gamma_s,\mu,\mu_0) \,= &  -\frac{\Gamma_s^0}{\beta_0}\,\left[
\ln r 
+  \left(\frac{\Gamma_s^1}{\Gamma_s^0}-\frac{\beta_1}{\beta_0}\right)
\frac{\alpha_s(\mu_0)}{4\pi}(r-1)\,
 + 
\left(\frac{\Gamma_s^2}{\Gamma_s^0}-\frac{\beta_2}{\beta_0}
  -\frac{\beta_1}{\beta_0}\left(\frac{\Gamma_s^1}{\Gamma_s^0}
  -\frac{\beta_1}{\beta_0}\right)\right)
  \frac{\alpha_s^2(\mu_0)}{2(4\pi)^2}\,\Big(r^2-1\Big)
+ \ldots
\right], 
\nn \\[1mm]
\tilde K(\Gamma_s,\gamma_s,\mu,\mu_0) \, = & \,\,\tilde
\omega\Big(\frac{\gamma_s}{2},\mu,\mu_0\Big) 
\, - \,\frac{\Gamma_s^0}{2\beta_0^2}\,
\bigg\{\,\frac{4\pi}{\alpha_s(\mu_0)}\,\left[\,1-\frac{1}{r}-\ln r\,\right]
+\,\left(\frac{\Gamma_s^1}{\Gamma_s^0}-\frac{\beta_1}{\beta_0}\right)
\left(1-r+\ln r\right)\, + \, \frac{\beta_1}{2\beta_0}\ln^2 r
\nn\\[1mm] & 
+ \, \left(\frac{\alpha_s(\mu_0)}{4\pi}\right)\,\bigg[\,
\bigg(\frac{\beta_1\Gamma_s^1}{\beta_0\Gamma_s^0}-\frac{\beta_2}{\beta_0}
       +\bigg(\frac{\beta_1\Gamma_s^1}{\beta_0\Gamma_s^0}-\frac{\beta_1^2}{\beta_0^2} 
     \bigg)(r-1)\bigg)\,\ln r
-\bigg( \frac{\beta_1\Gamma_s^1}{\beta_0\Gamma_s^0}-\frac{\beta_2}{\beta_0}\bigg)(r-1)
\nn\\[1mm] &  \qquad
+\bigg(\frac{\beta_1\Gamma_s^1}{\beta_0\Gamma_s^0}+\frac{\beta_2}{\beta_0}-\
   \frac{\Gamma_s^2}{\Gamma_s^0}-\frac{\beta_1^2}{\beta_0^2}\bigg)\frac{1}{2}(r-1)^2
\,\bigg]
\,\bigg\}\,,
\end{align}
where for $\tilde\omega$ the N${}^k$LL solutions correspond to the terms up to
order $\alpha_s^k$, and for $\tilde K$ the N${}^k$LL solutions
correspond to the terms up to order  $\alpha_s^{k-1}$. The solutions depend
on the cusp and the non-cusp anomalous dimensions of the soft function and the
beta-function [$\alpha_s=\alpha_s(\mu)$], 
\begin{align}
\label{cuspnoncusp1}
\Gamma_s[\,\alpha_s\,] & \, = \, -\Gamma_{\rm cusp}[\,\alpha_s\,] \, = \, 
\sum_{k=0}^\infty \,\,\left(\frac{\alpha_s}{4\pi}\right)^{k+1}\,\Gamma_s^{k}
\,,
\mbox{\hspace{2cm}}
\gamma_s[\,\alpha_s\,]  \, = \, 
\sum_{k=0}^\infty \,\,\left(\frac{\alpha_s}{4\pi}\right)^{k+1}\,\gamma_s^{k}
\,,
\nn\\[1mm]
\frac{d\alpha_s}{d\ln\mu} & \, = \,
\beta[\,\alpha_s\,] \, = \, - 2\alpha_s\,
\sum_{k=0}^\infty \,\,\left(\frac{\alpha_s}{4\pi}\right)^{k+1}\,\beta_{k}
\,,
\end{align}
The first few coefficients sufficient for NNLL order running read
\begin{align}
\label{Gammagamma}
\Gamma_s^0 \, = & \,-\,\Gamma_{\rm cusp}^0\,= \,-4 \,C_F
\,, 
\nn \\[1mm]
\Gamma_s^1 \, =  &\,-\,\Gamma_{\rm cusp}^1\,=  \,-4 \,C_A\,C_F\,
\left(\frac{67}{9} - \frac{\pi^2}{3}\right) + C_F\,T n_f \, \frac{80}{9}
\,,
\nn \\[1mm]
\Gamma_s^2 \, = & \,-\,\Gamma_{\rm cusp}^2\,=  
  C_A^2\,C_F\,\bigg(\!\!-\frac{490}{3} + \frac{536}{27}\pi^2 
         - \frac{44}{45}\pi^4 - \frac{88}{3}\zeta_3\bigg) 
+  C_A\,C_F\,T n_f\,\bigg(\frac{1672}{27} - \frac{160}{27}\pi^2 
         + \frac{224}{3}\zeta_3\bigg)
\nn \\ & \qquad\qquad
+ C_F^2\,T n_f\,\bigg(\frac{220}{3} - 64\zeta_3\bigg)
+\frac{64}{27}\,C_F\,(T n_f)^2
\,,
\nn \\[1mm]
\gamma_s^0 \, = & \,0
\,, 
\nn \\[1mm]
\gamma_s^1 \, =  & \, C_A\, C_F\, \left(-\frac{808}{27} + \frac{11}{9}\,\pi^2 
+ 28\,\zeta_3\right)
+C_F\,T n_f \left( \frac{224}{27} - \frac{4}{9}\, \pi^2 \right) 
\,.
\end{align}
and we also have 
\begin{align}
\beta_0 & \,=\, \frac{11}{3} C_A-\frac{4}{3} T n_f \,, 
\mbox{\hspace{3cm}}
\beta_1  \,=\, \frac{34}{3} C_A^2-\frac{20}{3} C_A\, T n_f - 4 C_F\, T n_f\,,
\nn\\[1mm]
\beta_2 &\,=\, \frac{2857}{54} C_A^3 - \frac{1415}{27} C_A^2 T n_f 
   + \frac{158}{27} C_A (T n_f)^2 
   + \frac{44}{9} C_F (T n_f)^2 - \frac{205}{9} C_A\, C_F\, T n_f + 2 C_F^2\,T n_f
\end{align}
for $n_f$ light flavors, from the running
of the strong coupling in the $\overline{\mbox{MS}}$ scheme.

It is an important feature of the factorization theorem that the RG properties
with respect to the variables $x_1$ and $x_2$ factorize such that in
$S(x_1,x_2,\mu)$ the $\mu$-dependence can only arise in terms of powers
of $\ln(x_{1,2}\mu)$. As shown in Ref.~\cite{Hoang:2007vb} it is therefore
possible to write Eq.~(\ref{UUS1}) in the form 
\begin{align}
\label{UUS2}
S(x_1,x_2,\mu) \, = \, 
U_s\left(x_1,\mu,(i\,x_1 e^{\gamma_E})^{-1}\right)\,
U_s\left(x_2,\mu,(i\,x_2 e^{\gamma_E})^{-1}\right)\,
\tilde S(x_1,x_2)\,, 
\end{align}
where $\tilde S$ is $\mu$-independent and 
$\tilde S(x_1,x_2)=\tilde S(x_2,x_1)$.
\\

\noindent
{\it Nonabelian Exponentiation.}
In Refs.~\cite{Gatheral:1983cz,Frenkel:1984pz} it has been proven to all orders
in perturbation theory that QCD matrix elements with arbitrary number of
external gluons exponentiate, if their operator definition can be written
entirely in terms of Wilson lines. If the external gluon final state is
symmetrized, the 
exponentiation also holds for contributions of these matrix elements to cross
sections and production rates. The exponentiation property also applies to the
hemisphere soft function and is most transparent in position space. Using
Eq.~(\ref{UUS2}) we can therefore write, to all orders in perturbation theory,
\begin{align}
\label{UUS3}
S_{\rm part}(x_1,x_2,\mu) \, = & \,\,
U_s\left(x_1,\mu,(i\,x_1 e^{\gamma_E})^{-1}\right)\,
U_s\left(x_2,\mu,(i\,x_2 e^{\gamma_E})^{-1}\right)\,
e^{T(x_1,x_2)}
\nn\\[1mm]
\, = & \,
\exp\left[
\tilde K\left(\Gamma_s,\gamma_s,\mu,(i\,x_1 e^{\gamma_E})^{-1}\right)
+ \tilde K\left(\Gamma_s,\gamma_s,\mu,(i\,x_2 e^{\gamma_E})^{-1}\right)
+T(x_1,x_2)\right]\,,
\end{align}
where $T(x_1,x_2)=T(x_2,x_1)$.
In this context exponentiation means that the argument of the exponential
has a simpler color structure than $S_{\rm part}(x_1,x_2,\mu)$ itself. There
are two specific features that are worth to mention:
(i) At ${\cal O}(\alpha_s^n)$ the highest power of logarithms of $x_{1,2}$ in
the exponent 
is $\ln^{n+1}$, while in $S$ it is $\ln^{2n}$ and (ii) the exponent does not
contain any $\alpha_s^n C_F^n$ terms except for $n=1$. The latter property
means that in QED the exponent is ${\cal O}(\alpha_s)$-exact and does not
contain any higher order corrections.
\\

\noindent
{\it Analytic form of $T(x_1,x_2)$.} It is now straightforward to determine
the analytic form of
the function $T(x_1,x_2)$ to ${\cal O}(\alpha_s^2)$. If there is
only a single gluon in the final state it is either in hemisphere~1 or in
hemisphere~2. Thus to ${\cal O}(\alpha_s)$ the soft function is a symmetric
sum of two terms 
each of which is either a function of $x_1$ or of $x_2$. If there are two
partons in the final state, they can be in different hemispheres, and a
non-trivial dependence on $x_1$ and $x_2$ can arise. Thus to ${\cal
O}(\alpha_s^2)$ the function $T$ must have the form
[$\mu_{x_{1,2}}  \equiv (i\,x_1 e^{\gamma_E})^{-1}$]
\begin{align}
\label{Tansatz}
T(x_1,x_2) \, = & \, 
\frac{\alpha_s\left(\mu_{x_1}\right)}{4\pi}\, t_1 +
\frac{\alpha_s\left(\mu_{x_2}\right)}{4\pi}\, t_1 +
2\,\frac{\alpha_s^2}{(4\pi)^2}\,t_2(x_1,x_2)\,,
\end{align}
where $t_1$ that can be read off from Eq.~(\ref{S1loop}),
\begin{align}
t_1 \, = \, -\,C_F\,\frac{\pi^2}{2}
\,.
\end{align} 
Moreover we have $t_2(x_1,x_2)=t_2(x_1/x_2)=t_2(x_2/x_1)$, since the $x_{1,2}$
are variables which have the dimension of an inverse mass. Here it is worth to
mention that the term $\alpha_s^2$ actually reads \
$c\,(\alpha_s^2(\mu_{x_{1}})+\alpha_s^2(\mu_{x_{2}}))+d\,\alpha_s(\mu_{x_{1}})\,
\alpha_s(\mu_{x_{2}})$ with $2\,c+d=1$. Since any event shape distribution
in the dijet limit
can be written in terms of delta functions and plus distributions, we can
further use the information that the momentum space soft function only depends
on $\ell^\pm$ through $\delta$-functions or plus-distributions. Thus in
position space $t_2$
can only depend on $x_{1,2}$ through even powers of $\ln(x_1/x_2)$, see
Eqs.~(\ref{FT3},\ref{FT4}). Using also
the constraint that $t_2$ cannot contain any term $\ln^n x_{1,2}$ with $n>
3$, it must have the simple form
\begin{align}
\label{t2defx}
t_2(x_1,x_2) \,=\, s_1\, + s_2\,\ln^2\Big(\frac{x_1}{x_2}\Big)
\,.
\end{align}
An important consequence of the exponentiation property is that $s_1$ and $s_2$ do
not contain any contribution proportional to the color factor $C_F^2$. This
feature serves as an important cross check for the numerical analysis
we carry out in Sec.~\ref{sectionevent2}.
\\

\noindent
{\it Form of the Soft Function.}
Expanded in $\alpha_s(\mu)$ and using the results collected above, the
${\cal O}(\alpha_s^2)$ position space hemisphere soft function reads
[$\alpha_s=\alpha_s(\mu), L_{1,2} \equiv \ln(i\,x_{1,2} e^{\gamma_E}\,\mu)$]
\begin{align}
\label{Sposition}
S_{\rm part} & ( x_1, x_2,\mu) \, = \, 
1 \, + \, \left(\frac{C_F\, \alpha_s}{4\pi}\right)\,
\bigg\{\!
-\bigg[4 L_1^2+\frac{\pi^2}{2}\bigg]
-\bigg[4 L_2^2+\frac{\pi^2}{2}\bigg]
\bigg\} 
\nn \\ &
+\left(\frac{\alpha_s}{4\pi}\right)^2\,\bigg\{ 
C_F^2\bigg[4 L_1^2+\frac{\pi^2}{2}\bigg]\bigg[4
L_2^2+\frac{\pi^2}{2}\bigg] 
\, + \, 8 C_F^2 (L_1^4+L_2^4)
\, + \, \bigg[\!-\frac{88}{9}C_A C_F+\frac{32}{9} C_F T n_f
\bigg](L_1^3+L_2^3)
\nn\\[1mm] & \mbox{\hspace{1cm}}
\, + \, \bigg[\, 2 C_F^2 \pi^2 
+ C_A C_F\bigg(-\frac{268}{9}+\frac{4}{3}\pi^2\bigg)
+ \frac{80}{9} C_F T n_f \bigg](L_1^2+L_2^2)
\nn\\[1mm] & \mbox{\hspace{1cm}}
\, + \, \bigg[\, 
C_A C_F\bigg(-\frac{808}{27}-\frac{22}{9}\pi^2+28 \zeta_3 \bigg)
+ C_F T n_f\bigg(\frac{224}{27}+\frac{8}{9}\pi^2\bigg)
\bigg](L_1+L_2)
\nn\\[1mm] & \mbox{\hspace{1cm}}
\, + \, \frac{\pi^4}{4}C_F^2 + 2\, t_2(x_1,x_2)
\bigg\}\,.
\end{align}
The corresponding momentum space hemisphere soft function has the form 
[${\cal L}_\pm^n\equiv 1/\mu \big[\theta(\ell^\pm)\ln^n(\ell^\pm/\mu)/(\ell^\pm/\mu)\big]_+$]
\begin{align}
\label{Smomentum}
S_{\rm part} & (\ell^+,\ell^-,\mu) \, = \,
\delta(\ell^+)\,\delta(\ell^-)
\nn\\[1mm] & 
+  \left(\frac{C_F\,\alpha_s}{4\pi}\right)\,
\bigg\{
\bigg[\!-8{\cal L}_+^1 + \frac{\pi^2}{6}\,\delta(\ell^+)\bigg]\,\delta(\ell^-) 
 + 
\bigg[\!-8{\cal L}_-^1 + \frac{\pi^2}{6}\,\delta(\ell^-)\bigg]\,\delta(\ell^+) 
\bigg\}
\nn\\[1mm] & 
+  \left(\frac{\alpha_s}{4\pi}\right)^2\,\bigg\{\,
C_F^2 \bigg[\!-8{\cal L}_+^1 +\frac{\pi^2}{6}\,\delta(\ell^+)\bigg]
\bigg[\!-8{\cal L}_-^1+\frac{\pi^2}{6}\,\delta(\ell^-)\bigg]
+ 32 C_F^2
\Big[  \delta(\ell^+){\cal L}_-^3 + \delta(\ell^-){\cal L}_+^3 \Big]
\nn\\[1mm] &  \mbox{\hspace{0.5cm}}
+ \bigg[\,\frac{88}{3}C_A C_F-\frac{32}{3}C_F T n_f\bigg]
\Big[  \delta(\ell^+){\cal L}_-^2 + \delta(\ell^-){\cal L}_+^2 \Big]
\nn\\[1mm] &  \mbox{\hspace{0.5cm}}
+ \bigg[\!-12\pi^2 C_F^2
  +C_A C_F \bigg(\!-\frac{536}{9}+\frac{8}{3}\pi^2\bigg)
  +\frac{160}{9} C_F Tn_f\bigg]
\Big[  \delta(\ell^+){\cal L}_-^1 + \delta(\ell^-){\cal L}_+^1 \Big]
\nn\\[1mm] &  \mbox{\hspace{0.5cm}}
+ \bigg[\, 64 \zeta_3 C_F^2 
  +C_A C_F \bigg(\frac{808}{27}-\frac{22}{9}\pi^2-28\zeta_3\bigg)
  +C_F Tn_f\bigg(\!-\frac{224}{27}+\frac{8}{9}\pi^2\bigg)\bigg]
\Big[  \delta(\ell^+){\cal L}_-^0 + \delta(\ell^-){\cal L}_+^0 \Big]
\nn\\[1mm] &  \mbox{\hspace{0.5cm}}
+ \bigg[\!-\frac{3}{40}\pi^4 C_F^2 
  +C_A C_F \bigg(\frac{134}{27}\pi^2-\frac{2}{9}\pi^4+\frac{176}{9}\zeta_3\bigg)
  +C_F Tn_f\bigg(\!-\frac{40}{27}\pi^2-\frac{64}{9}\zeta_3\bigg)\bigg]
\,2\,\delta(\ell^+)\delta(\ell^-) 
\nn\\[1mm] &  \mbox{\hspace{0.5cm}}
\, +\,  2\,t_2(\ell^+,\ell^-)
\,\bigg\}\,,
\end{align}
where
\begin{align}
\label{t2defmom}
t_2(\ell^+,\ell^-) &\, = \,
s_1\,\delta(\ell^+)\,\delta(\ell^-) 
\, + \,
s_2\,\bigg\{\,
\bigg[ 2{\cal L}_+^1-\frac{\pi^2}{6}\delta(\ell^+) \bigg]\,\delta(\ell^-) 
\, + \,
\bigg[ 2{\cal L}_-^1-\frac{\pi^2}{6}\delta(\ell^-) \bigg]\,\delta(\ell^+) 
-2\,{\cal L}_+^0\,{\cal L}_-^0
\,\bigg\}
\,.
\end{align}
Note that while the distributions ${\cal L}_\pm^n$ that appear in
Eq.~(\ref{t2defmom}) depend on the renormalization scale $\mu$, this
$\mu$-dependence cancels in the combination of all terms. 
For completeness we also present the RG evolution of the 
momentum space soft function. It has the form
$S(\ell^+,\ell^-,\mu) = \int d\ell^{\prime +}\ell^{\prime -} 
U_s(\ell^+-\ell^{\prime +},\mu,\mu_0)U_s(\ell^--\ell^{\prime
-},\mu,\mu_0)S(\ell^{\prime +},\ell^{\prime -},\mu_0)$ with
\begin{align}
\label{Usdefmom}
U_s(\ell,\mu,\mu_0) \, = \,
\frac{e^{\tilde K(\Gamma_s\gamma_s,\mu,\mu_0)}(e^{\gamma_E})^{\tilde\omega(\Gamma_s,\mu,\mu_0)}}
     {\mu_0\,\Gamma(-\tilde\omega(\Gamma_s,\mu,\mu_0))}\,
 \bigg[ 
\frac{(\mu_0)^{1+\tilde\omega(\Gamma_s,\mu,\mu_0)}\,\theta(\ell)}{\ell^{1+\tilde\omega(\Gamma_s,\mu,\mu_0)}}
 \bigg]_+
\,,
\end{align}
where $\tilde\omega$ and $\tilde K$ are given in Eqs.~(\ref{omegaK}). The
result is obtained without any effort using the position space result in
Eq.~(\ref{Usdefx}) and the Fourier transformation given in Eq.~(\ref{FT3}).
The plus function  with an arbitrary exponent $1+\tilde \omega$ with
$\tilde\omega<1$ is defined by
\begin{align}
\label{plusfctdef}
  \bigg[ \frac{\theta(x)}{(x)^{1+\tilde\omega}}\bigg]_+ &=  \lim_{\beta\to 0} \bigg[
  \frac{\theta(x\minus \beta)}{(x)^{1+\omega}} -
   \delta(x\minus \beta) \: \frac{\beta^{-\tilde\omega}}{\tilde\omega} \bigg]
\,.
\end{align}

\section{Thrust and Heavy Jet Mass Distributions}
\label{sectiondistribution}

In this section we determine the ${\cal O}(\alpha_s^2)$ fixed-order cumulative
distributions for the thrust and the heavy jet mass variables, and for their
generalizations, called $\tau_\alpha$ and $\rho_\alpha$ and defined
below. The distributions are derived from the dijet factorization
theorem~(\ref{facttheoSCET}) for the double differential hemisphere invariant
mass distribution for massless quark production. At ${\cal O}(\alpha_s^2)$ the
distributions depend on the 
parameters $s_1$ and $s_2$ in the hemisphere soft function that are determined
numerically in Sec.~\ref{sectionevent2}. Since EVENT2 produces ${\cal
O}(\alpha_s^2)$ distributions for the renormalization scale $\mu=Q$ we also
use this scale choice for all the evaluations that follow in this section. This
means in particular that any summation of logarithms in the hard
coefficient, the jet function or soft function is neglected.
Since the renormalization scale is fixed to $\mu=Q$ it is convenient to use
the dimensionless variables $\tilde x_{1,2}$ and
\begin{align}
\tilde M_{1,2}^2 \, \equiv \, \frac{M_{1,2}^2}{Q^2}\,,&&
\tilde \ell^\pm \, \equiv \, \frac{\ell^\pm}{Q}\,.
\end{align}

We start by writing down the double differential hemisphere invariant mass
spectrum in the dijet region shown in Eq.~(\ref{facttheoSCET}) in position
space representation normalized to 
the tree level total cross section, 
\begin{align}
\label{facttheoSCETposition}
\sigma(\tilde x_1,\tilde x_2)& \, \equiv \,
\frac{1}{\sigma_0}\frac{d\sigma^{\rm dijet}}{d\tilde x_1\, d\tilde x_2} \, = \,
\int d \tilde M_{1}^2\, d\tilde M_{2}^2
\, e^{-i \tilde M_{1}^2\tilde x_1}
\, e^{-i \tilde M_{2}^2\tilde x_2}\,
\frac{1}{\sigma_0}
\frac{d\sigma^{\rm dijet}}{d\tilde M_{1}^2\, d\tilde M_{2}^2}
\nn\\[1mm] &
\, = \,
H_Q(Q,Q) \, J_1(\tilde x_1/Q^2,Q)\, J_2(\tilde x_2/Q^2,Q)
\, S(\tilde x_1/Q,\tilde x_2/Q,Q)\,.
\end{align}
We now define the event-shape variables $\tau_\alpha$ and $\rho_\alpha$ as
\begin{align}
\label{taurhodef}
\tau_\alpha \, \equiv \, & \,\frac{2}{1+\alpha}\,\frac{\alpha \, M_1^2+M_2^2}{Q^2} 
\, = \, \frac{2\,(\alpha\,\tilde M_1^2+\tilde M_2^2)}{1+\alpha}\,,
\nn\\[1mm]
\rho_\alpha \, \equiv \, & \,\frac{2}{1+\alpha}\,\frac{1}{Q^2}\,{\rm max}(\alpha \, M_1^2,M_2^2) 
\, = \, \frac{2}{1+\alpha}\,{\rm max}(\alpha\,\tilde M_1^2,\tilde M_2^2)\,.
\end{align}
For $\alpha=1$, $\tau_\alpha$ and $\rho_\alpha$ reduce to the common thrust
and heavy jet mass variables. For $\alpha\neq 1$ the two hemispheres get
different weights and probe the $M_1^2$-$M_2^2$ distribution
asymmetrically. From the $M_1^2\leftrightarrow M_2^2$ symmetry one can derive
the relations
\begin{align}
\label{alphasymmetry}
\tau_\alpha \, = \, \tau_{1/\alpha}\,, 
\mbox{\hspace{1cm}}\mbox{and}
\mbox{\hspace{1cm}}
\rho_\alpha \, = \, \rho_{1/\alpha}\,.
\end{align}
We note that, if $\alpha$ is chosen much larger or smaller than one, the
$\tau_\alpha$ and $\rho_\alpha$  distributions in fixed-order perturbation
theory develop large logarithms of 
$\alpha$. These logarithms are examples of non-global logarithms~\cite{Dasgupta:2001sh},
and they can be summed in the factorization theorem by an
independent setting of the renormalization scales that govern the invariant
masses of the two hemispheres.
It is now straightforward to determine the ${\cal O}(\alpha_s^2)$
fixed-order cumulative $\tau_\alpha$ and $\rho_\alpha$ distributions
in the dijet limit for $\mu=Q$ 
\begin{align}
\label{Sigmathrust}
\Sigma_{\tau_\alpha}^{\rm dijet}(\Omega) \,& = \, 
\int_0^\Omega \!\! d\tau_\alpha\,\frac{1}{\sigma_0}\,
\frac{d\sigma^{\rm dijet}}{d\tau_\alpha}  = 
\int_0^\Omega \!\! d\tau_\alpha\,
d\tilde M_1^2\, d\tilde M_2^2\,
\delta(\tau_\alpha-{\textstyle\frac{2}{1+\alpha}}(\alpha \tilde M_1^2+\tilde M_2^2))
\,\frac{1}{\sigma_0}\,
\frac{d\sigma^{\rm dijet}}{d\tilde M_1^2 d\tilde M_2^2}
\nn\\[1mm] & = \,-\,i\,\int_{-\infty}^{+\infty}\!\!\frac{d\tilde x}{2\pi}\,\,
e^{i\,\Omega\,\tilde x}\,\,
\frac{\sigma(\alpha\tilde x,\tilde x)}{\tilde x-i 0}
\,,
\\[1mm]
\label{Sigmarho}
\Sigma_{\rho_\alpha}^{\rm dijet}(\Omega) \,& = \, 
\int_0^\Omega \!\! d\rho_\alpha\,\frac{1}{\sigma_0}\,
\frac{d\sigma^{\rm dijet}}{d\rho_\alpha}  = 
\int_0^\Omega \!\! d\rho_\alpha\,
d\tilde M_1^2\, d\tilde M_2^2\,
\delta(\rho_\alpha-{\textstyle\frac{2}{1+\alpha}}{\rm max}(\alpha \tilde M_1^2,\tilde M_2^2))
\,\frac{1}{\sigma_0}\,
\frac{d\sigma^{\rm dijet}}{d\tilde M_1^2 d\tilde M_2^2}
\nn\\[1mm] & = \,-\,\int_{-\infty}^{+\infty}\!\!
\frac{d\tilde x_1}{2\pi}\,
\frac{d\tilde x_2}{2\pi}\,\,
e^{i\,\Omega\,(\tilde x_1+\tilde x_2)}\,\,
\frac{\sigma(\alpha\tilde x_1,\tilde x_2)}{(\tilde x_1-i 0)(\tilde x_2-i 0)}
\,.
\end{align}
The analytic expressions for the cumulative distributions above are given in
App.~\ref{app:distributions}. Note that the function $t_2$ in
Eqs.~(\ref{Sposition}) and (\ref{Smomentum}) does not lead to any logarithmic
terms involving the cut-off $\Omega$ in the cumulative distributions. 
While in one dimension such a behavior can, in momentum space, only be
obtained from a delta function, in several dimensions, it can also be achieved
by proper combinations of plus-distributions as shown in Eq.~(\ref{t2defmom}).
This behavior is easier to see in position space where $t_2$ only depends on
the ratio $\tilde x_1/\tilde x_2$ and contributes as  
\begin{align}
\label{t2explicit}
t_{2,\tau_\alpha} \, & = \,-\,i\,\int_{-\infty}^{+\infty}\!\!\frac{d\tilde x}{2\pi}\,\,
e^{i\,\Omega\,\tilde x}\,\,
\frac{t_2(\alpha)}{\tilde x-i 0} 
\, = \, \theta(\Omega)\,\Big[\,s_1 \, + \, s_2\,\ln^2\alpha\,\Big]\,,
\nn\\[1mm]
t_{2,\rho_\alpha} \, & = \,-\,\int_{-\infty}^{+\infty}\!\!
\frac{d\tilde x_1}{2\pi}\,
\frac{d\tilde x_2}{2\pi}\,\,
e^{i\,\Omega\,(\tilde x_1+\tilde x_2)}\,\,
\frac{t_2(\alpha\tilde x_1/\tilde x_2)}{(\tilde x_1-i 0)(\tilde x_2-i 0)}
\, = \,
\theta(\Omega)\,\Big[\,s_1 \, + \, s_2\,\Big(\ln^2\alpha-\frac{\pi^2}{3}\Big)\,\Big]\,,
\end{align}
in Eqs.~(\ref{Sigmathrust}) and (\ref{Sigmarho}). The constants depend on
the  definition of the event-shape variable and in particular on the value of
$\alpha$.

\newpage
\section{Numerical Analysis using EVENT2}
\label{sectionevent2}

In this section we determine the constants $s_1$ and $s_2$ in Eqs.~(\ref{t2defx}),
(\ref{t2defmom}), which cannot be determined from the general
arguments discussed in Sec.~\ref{sectionproperties}. 
\\

\noindent
{\it Method.}
The EVENT2 program determines numerical estimates for event-shape
distributions at ${\cal O}(\alpha_s^2)$ in the fixed-order expansion for
$\mu=Q$ in full QCD. Using the variable $y$ generically for $\tau$,
$\tau_\alpha$ or $\rho_\alpha$, the distributions have the form
\begin{align}
\frac{1}{\sigma_0}\,\frac{d\sigma}{dy} \, = \,
A(y)\, + \,\
\left(\frac{\alpha_s(Q)}{2\pi}\right)\,B(y) \, + \,
\left(\frac{\alpha_s(Q)}{2\pi}\right)^2\,C(y)\, + \, \ldots
\,.
\end{align}
The corresponding cumulative distributions read
\begin{align}
\Sigma_y(\Omega) \, = \,
\int_0^\Omega\!\! dy\,\frac{1}{\sigma_0}\,\frac{d\sigma}{dy} \, = \,
\Sigma_y^{(0)}(\Omega) \, + \,
\left(\frac{\alpha_s(Q)}{2\pi}\right)\,\Sigma_y^{(1)}(\Omega) \, + \,
\left(\frac{\alpha_s(Q)}{2\pi}\right)^2\,\Sigma_y^{(2)}(\Omega) \, + \,\ldots\,.
\end{align}
Defining bin boundaries $\Omega^n$ for $n=0,1,\ldots,n_{\rm max}$ with
$0<\Omega^n <\Omega^{n+1}$ and $\Omega^{n_{\rm max}}=1$, EVENT2 can determine
the sum of weights of events falling into the $n_{\rm max}$ bins, which for
the $n$th bin represents a numerical estimate for 
\begin{align}
\Delta\sigma_n \,&  \equiv \, 
\Sigma_y(\Omega^n) \, - \, \Sigma_y(\Omega^{n-1}) \, = \,
\int_{\Omega^{n-1}}^{\Omega^{n}}\! dy\,\frac{1}{\sigma_0}\,\frac{d\sigma}{dy}
\nn\\[1mm] & =  \,
\Delta\sigma_n^{(0)} \, + \,
\left(\frac{\alpha_s(Q)}{2\pi}\right)\,\Delta\sigma_n^{(1)} \, + \,
\left(\frac{\alpha_s(Q)}{2\pi}\right)^2\,\Delta\sigma_n^{(2)} \, + \, \ldots
\,.
\end{align} 
To determine the unknown constants $s_1$ and $s_2$ in the dijet distribution
from Eq.~(\ref{facttheoSCET}), one can use the fact that they do not appear in 
$\Delta\sigma_n$ for any choice of bin boundaries. This is because a
dependence on $s_{1,2}$ can only appear in integrations that
contain the threshold at 
$y=0$, see Eqs.~(\ref{Sigmathrust}), (\ref{Sigmarho}) and
(\ref{t2explicit}). The method starts by subtracting the known 
dijet contributions $\Delta\sigma_n^{(0),{\rm dijet}}$  from the full theory
$\Delta\sigma_n^{(2)}$ computed by EVENT2. For $d\sigma/dy$ the difference is
at most logarithmically singular for $y\to 0$ and thus integrable at
$y=0$. Thus for the cumulative distribution the remainder
\begin{align}
\label{sigmadiff}
\Sigma_y^{\rm rest}(\Omega)\, \equiv \,
\Sigma_y(\Omega)\, - \, \Sigma_y^{\rm dijet}(\Omega)
\end{align}
vanishes for $\Omega\to 0$. The remainder distribution is also independent of
$s_1$ and $s_2$ for any $\Omega>0$. It is the aim to determine a numerical
estimate for $\Sigma^{(2),{\rm rest}}(1)$ from EVENT2. Using that
$\Sigma_y(1)$ is equal to the total cross
section~\cite{Chetyrkin:1979bj,Dine:1979qh,Celmaster:1979xr}, 
\begin{align}
\Sigma_y(1) \,& = \, R(Q^2) \, = \, \frac{\sigma_{\rm tot}}{\sigma_0}
\, = \,
1 \, + \,
\frac{3}{2}\,\left(\frac{C_F\, \alpha_s(Q)}{2\pi}\right) \, + \,
\left(\frac{\alpha_s(Q)}{2\pi}\right)^2\, r_2 \, + \, \ldots \,,
\nn \\[2mm]
r_2 \, & = \, -\,\frac{3}{8}\,C_F^2 \, + \,
C_A\,C_F\,\bigg[\frac{123}{8}-11\,\zeta_3\bigg] \, + \,
C_F\,T n_f\,\bigg[\!\!
-\frac{11}{2}+\frac{1}{4}\,\zeta_3
\bigg]\,,
\end{align}
we can then determine the constants $t_{2,y}$ for the different event-shapes
using Eqs.~(\ref{taualphacumulativ}) and (\ref{rhoalphacumulativ}) and the
relation
\begin{align}
\label{sigmarelation}
\Sigma_y^{(2),{\rm dijet}}(1) \, = \, r_2 \, - \, \Sigma_y^{(2),{\rm rest}}(1)
\,.
\end{align} 

\vskip 3mm
\noindent
{\it Sum of all Color Factors.}
For our numerical analysis we ran EVENT2 with $10^{10}$ events  for $n_f=4$
and for $n_f=5$.\footnote{The task was distributed over 100
parallel jobs and took less than 30 hours to complete for each value
of $n_f$.}
We used $80$ bins from $10^{-4}$ to $1$ with
logarithmic bin boundaries located at $\Omega_n=10^{(n-80)/20}$ with
$n=0,1,\ldots,80$. Since EVENT2 works with an internal infrared
cut-off (set to $10^{-8}$), it is
not possible to obtain 
numerical results for a bin with a lower boundary located at $\Omega=0$. Thus to
obtain a numerical estimate for $\Sigma_y^{(2),{\rm rest}}(1)$ one has to
rely on an extrapolation of the numerical results for $\Sigma_y^{(2),{\rm
rest}}(1)-\Sigma_y^{(2),{\rm rest}}(\Omega)$ taking the limit $\Omega\to 0$. 

\begin{figure}[t]
\begin{center}
\epsfxsize=\textwidth
\epsffile{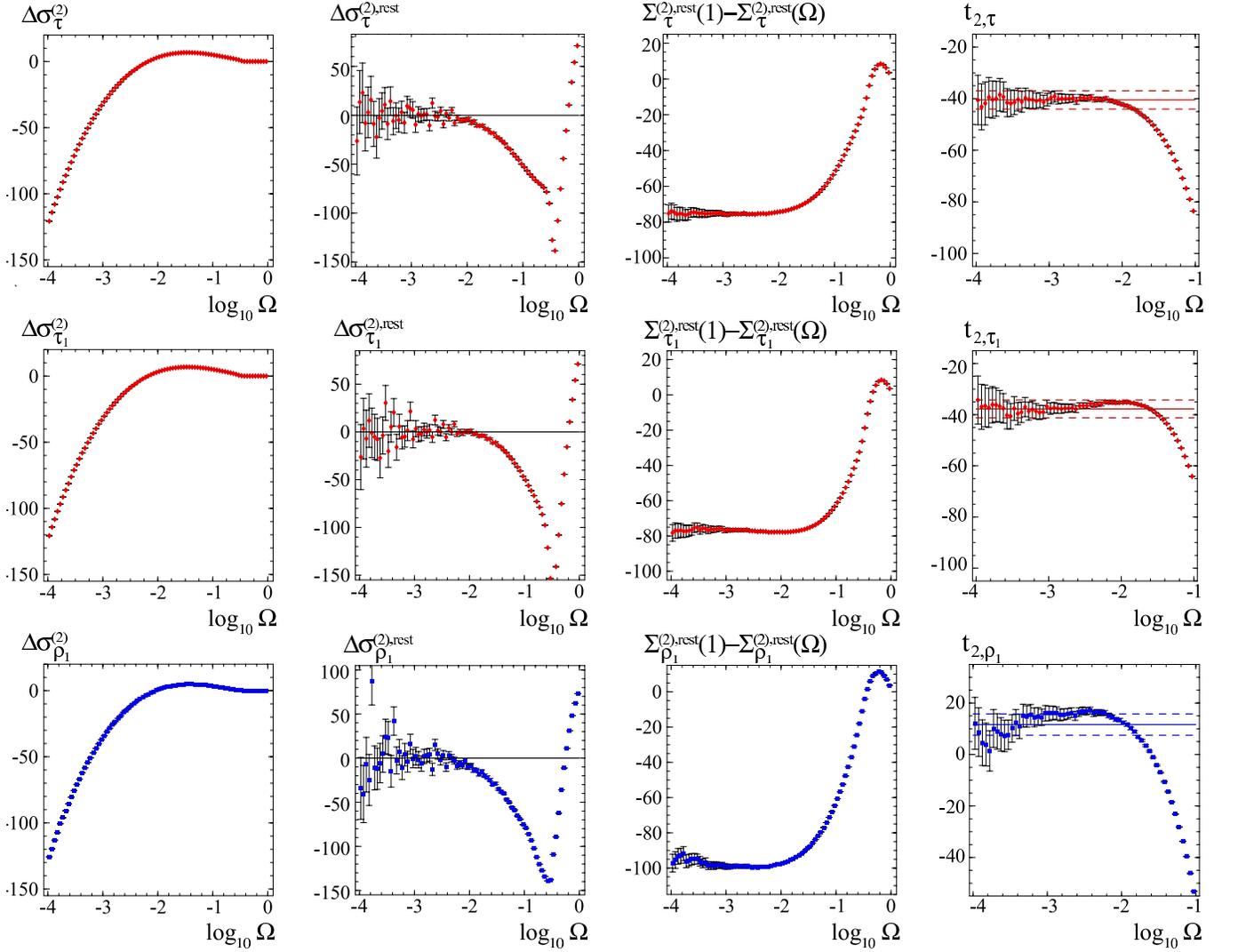}
\vskip -0.1cm
\caption{ 
Results of the numerical analysis to obtain $\Sigma_y^{(2),{\rm rest}}(1)$ and
$t_{2,y}$ for $y=\tau$ (top), $\tau_1$ (middle) and $\rho_1$ (lower row) for
$n_f=5$. The first column shows the full QCD results for
the $y$-distributions in bins with center at $\Omega$ obtained from EVENT2.
The second column shows the QCD distributions minus the singular
dijet contributions described by the factorization theorem~(\ref{facttheoSCET}).
The third column shows  $\Sigma_y^{(2),{\rm
    rest}}(1)-\Sigma_y^{(2),{\rm rest}}(\Omega)$, and the last column
shows $t_{2,y}$, taking  $\Sigma_y^{(2),{\rm
    rest}}(1)-\Sigma_y^{(2),{\rm rest}}(\Omega)$ as an estimate for 
$\Sigma_y^{(2),{\rm rest}}(1)$. The solid and dashed lines in the
panels in the column on the right 
represent the central value and the errors for $t_{2,y}$ which are
estimated from the limit of small $\Omega$.
 }
\label{fig:aeq1}
\end{center}
\end{figure}

Figure~\ref{fig:aeq1} shows the results of our numerical analysis to obtain
estimates for $\Sigma_y^{(2),{\rm rest}}(1)$ for the variables $y=\tau$,
$\tau_1$ and $\rho_1$ for $n_f=5$.\footnote{Our results for $n_f=4$ are
analogous and not discussed in detail.} The first row is for $\tau$, the
middle row for $\tau_1$ and bottom row for $\rho_1$.
We note that the thrust $\tau$ and the variable $\tau_1$ 
agree in the dijet limit, but they differ concerning power corrections and in
the multi-jet region where $\tau\sim\tau_1\sim {\cal O}(1)$. The
panels in the first column show the binned full QCD distributions
$\Delta\sigma_{y,n}^{(2)}$ as obtained from EVENT2. The central value
for each bin is displayed as a (colored) symbol 
and the statistical error as a vertical line. The panels in the second column
show the remainder distribution $\Delta\sigma_{y,n}^{(2),{\rm rest}}$ after
the singular dijet contributions have been subtracted. We see that the
remainder distribution falls off to zero for small $\Omega$ and that the
asymptotic regime $\Omega\to 0$ appears to be reached already for $\Omega\lsim
10^{-2.5}$. The statistical errors grow for decreasing $\Omega$ for the
remainder distribution because EVENT2 attempts to obtain a constant relative
statistical error for a given binning in the full QCD distributions shown in
the first column, and because the contributions from the singular dijet
terms increasingly dominate in size for $\Omega\to 0$. 
The panels in the third column show the integral over the
remainder distribution $\Sigma_y^{(2),{\rm rest}}(1)-\Sigma_y^{(2),{\rm
rest}}(\Omega)$, which has to be extrapolated for $\Omega\to 0$. Finally, the
panels in the column on the right display the estimates for the
constants $t_{2,y}$ adopting  $\Sigma_y^{(2),{\rm
    rest}}(1)-\Sigma_y^{(2),{\rm rest}}(\Omega)$ as shown in respective
panels in the third column  
for $\Sigma_y^{(2),{\rm rest}}(1)$ in Eqs.~(\ref{sigmarelation}). 
We use the average of the central values
for $\log_{10}\Omega<-2.5$ to estimate our final numbers for  $t_{2,y}$ and adopt
the error at $\log_{10}\Omega\approx -3.25$ as the uncertainty. The results of the
estimate are illustrated by the solid and dashed horizontal lines
in the panels in the last column. 
We note that our results are fully compatible with the theoretical expectation
that $t_{2,\tau}=t_{2,\tau_1}$. Using the relations
\begin{align}
s_1 \, = \, \frac{1}{2}\left(t_{2,\tau}+t_{2,\tau_1}\right)\,, & &
s_2 \, = \, \frac{3}{\pi^2}\,
\Big[\frac{1}{2}\left(t_{2,\tau}+t_{2,\tau_1}\right)- t_{2,\rho_1}\Big]\,,
\end{align}
we obtain
\begin{align}
\label{abfinal}
s_1 \, = \, \left\{\begin{array}{lr} -39.1 \pm 2.5 & (n_f=5) \\ 
-53.3 \pm 2.5 & (n_f=4)\end{array}\right.
\,, &&
s_2 \, = \, \left\{\begin{array}{lr} -15.4 \pm 1.5 & (n_f=5) \\ 
-14.9 \pm 1.5 &  (n_f=4)\end{array}\right.
\,.
\end{align} 
A method similar to ours has been applied recently in
Ref.~\cite{Becher:2008cf} to determine ${\cal O}(\alpha_s^2)$ corrections to
the soft function for thrust given in Eq.~(\ref{Sthrust}). The thrust
soft function depends on $s_1$, but has no dependence on $s_2$, see
Eq.~(\ref{t2explicit}). 
Transferred to our notation their results reads
$s_1(n_f=5)=-40.1\pm 3.1$ and $s_1(n_f=4)=-54.4\pm 3.0$. 
The results are compatible to ours.
In Ref.~\cite{Becher:2008cf} $10^{10}$ events were used,
but with linear binning and for $\Omega\ge 10^{-3}$.

\begin{figure}[t]
\begin{center}
\epsfxsize=9cm
\epsffile{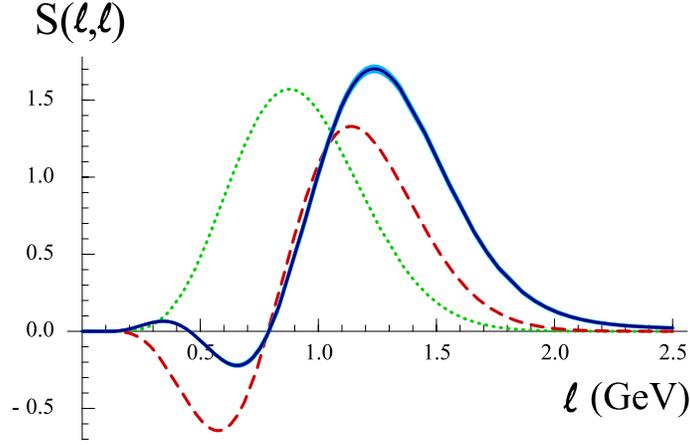}
\vskip 0.0cm
\caption{
Tree-level (green dotted line), one-loop (red dashed line) and two-loop (blue
solid line) soft function $S(\ell,\ell,\mu)$ (see Eq.~(\ref{Sdef2}))  without
renormalon subtraction for $\mu=1.5$~GeV ($\alpha_s(1.5~\mbox{GeV})=0.3285$)
and $n_f=5$. For the model function the parameters $\Lambda=0.55$~GeV,
$\Delta=0.1$~GeV  and $(a,b)=(3.0,-0.5)$ are used.
 }
\label{fig:softpure}
\end{center}
\end{figure}

In Fig.~\ref{fig:softpure} the tree-level (green dotted line), ${\cal
O}(\alpha_s)$ (red dashed line) and ${\cal O}(\alpha_s^2)$ (blue solid line)
soft function $S(\ell^+,\ell^-,\mu)$ including the convolution with the model
function as defined in Eq.~(\ref{Sdef2}) but without renormalon subtraction
are plotted over $\ell=\ell^+=\ell^-$ 
for $\mu=1.5$~GeV ($\alpha_s(1.5~\mbox{GeV})=0.3285$) and $n_f=5$, and using
$\Lambda=0.55$~GeV, $\Delta=0.1$~GeV  and $(a,b)=(3.0,-0.5)$ for the model
function in Eqs.~(\ref{fmod1}) and (\ref{Sgap}). For the partonic ${\cal
O}(\alpha_s^2)$ soft function we have adopted the results in
Eqs.~(\ref{abfinal}) for the constants $s_{1,2}$. The corresponding uncertainty
in the ${\cal O}(\alpha_s^2)$ soft function is visualized by the additional
light blue solid lines. The uncertainty is at the percent level where the soft
function is large and absolutely negligible in comparison to the remaining
perturbative QCD uncertainties.. We will therefore adopt the
central values given in Eqs.~(\ref{abfinal}) from now on without further
discussion. The rather poor perturbative behavior of the curves shown in
Fig.~\ref{fig:softpure} with the unphysical negative values and the
significant changes in the shape at higher orders is symptomatic for any
choice of model parameters and renormalization scale and illustrates
necessity to introduce the renormalon-free gap parameter. This will be
discussed in Sec.~\ref{sectiongap}.

\vskip 3mm
\noindent
{\it Cross Check using $\tau_\alpha$ and $\rho_\alpha$ for $\alpha\neq 1$.}
An important cross check of the results in Eqs.~(\ref{abfinal}) and also of
the form for the function $t_2$ in Eqs.~(\ref{t2defx}) and (\ref{t2defmom}) is
provided by comparing predictions for $t_{2,\tau_\alpha}$ and
$t_{2,\rho_\alpha}$ for $\alpha\neq 1$ with the corresponding
results obtained from EVENT2 based on a numerical analysis analogous to
the one in the previous section. 

\begin{figure}[t]
\begin{center}
\epsfxsize=\textwidth
\epsffile{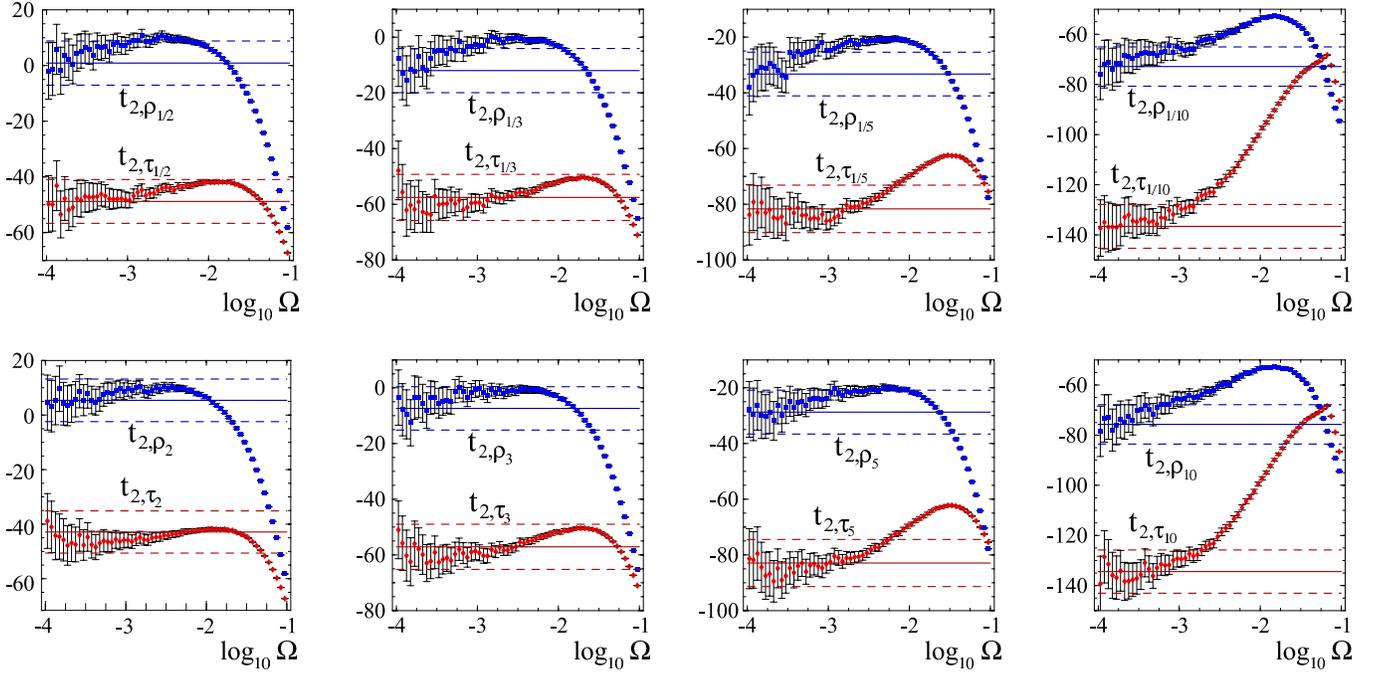}
\vskip -0.2cm
\caption{
Analysis and results with errors
for $t_{2,y}$ for $y=\tau_\alpha$ (blue symbols and lines)
and $y=\rho_\alpha$ (red symbols and lines) with
$\alpha=1,2,3,5,10$ and $1/2,1/3,1/5,1/10$ and using $n_f=5$.
 }
\label{fig:anot1}
\end{center}
\end{figure}

\begin{figure}[t]
\begin{center}
\epsfxsize=10cm
\epsffile{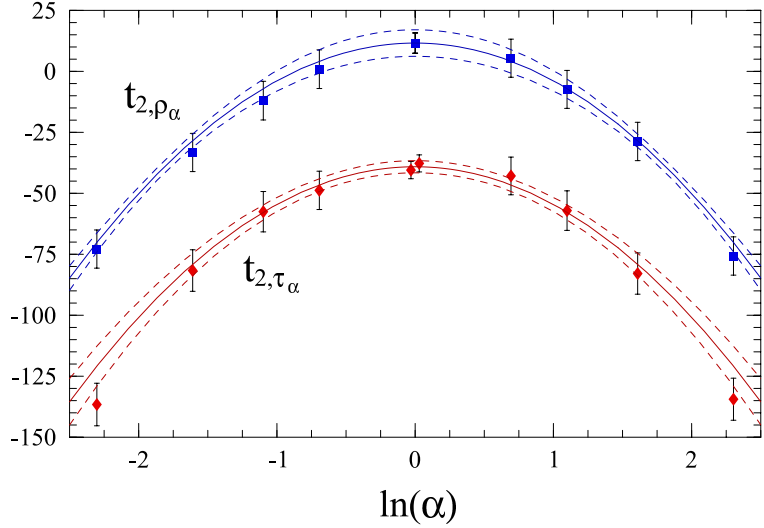}
\vskip -0.3cm
\caption{ 
Estimates for $t_{2,y}$ with uncertainties for $y=\tau_\alpha$ (lower red dots with
error bars) and $y=\rho_\alpha$ (upper blue dots with error bars) obtained
from EVENT2. The solid and dashed lines represent the predictions for
$t_{2,y}$ with uncertainties based on the results in Eqs.~(\ref{abfinal})
which are obtained from $t_{2,y}$ for $y=\tau,\tau_1,\rho_1$ (dots with error
bars at $\ln\alpha=0$). All results are for $n_f=5$.
 }
\label{fig:fulla}
\end{center}
\end{figure}

The results for $t_{2,\tau_\alpha}$ and $t_{2,\rho_\alpha}$ with
$\alpha=2,3,5,10$ and $1/2,1/3,1/5,1/10$ for $n_f=5$ are 
displayed in Fig.~\ref{fig:anot1}, where we use the type of
presentation from the last column in Fig.~\ref{fig:aeq1}. Comparing to
the results displayed in Fig.~\ref{fig:aeq1} 
we see that for increasing values of $\ln^2(\alpha)$ the asymptotic regime is
shifted towards smaller values of $\Omega$. Thus for estimating the values for 
$t_{2,\tau_\alpha}$ and $t_{2,\rho_\alpha}$ we now use the average of the lowest
five bins with $\log_{10}\Omega\le -3.75$ and adopt the error at $\log_{10}\Omega=-3.75$ as
the uncertainty. The results are displayed as horizontal lines in
Fig.~\ref{fig:anot1} and also summarized in Fig.~\ref{fig:fulla}. In
Fig.~\ref{fig:fulla} we have also shown
the theoretical predictions for $t_{2,\tau_\alpha}$ and $t_{2,\rho_\alpha}$
based on Eqs.~(\ref{t2explicit}) and the values of 
$s_1$ and $s_2$ from Eq.(\ref{abfinal}) with $n_f=5$ for their respective central
values (solid lines) and for their one-standard deviations (dashed
lines). The agreement is excellent and reassures the form of
Eq.~(\ref{t2explicit}). 
\\

\noindent
{\it Contributions from Different Color Factors.}
EVENT2 can give the ${\cal O}(\alpha_s^2)$ contributions for the distributions
separated with respect to the color factors $C_F^2, C_A C_F$ and $C_F T n_f$
and it is thus straightforward to determine the color factor components of the
constants $s_1$ and $s_2$. The numerical determination of the $C_F^2$
contributions is particular important since due to the nonabelian
exponentiation property of the soft function $s_1$ and $s_2$ do not have any
term proportional to $C_F^2$.

\begin{figure}[t]
\begin{center}
\epsfxsize=\textwidth
\epsffile{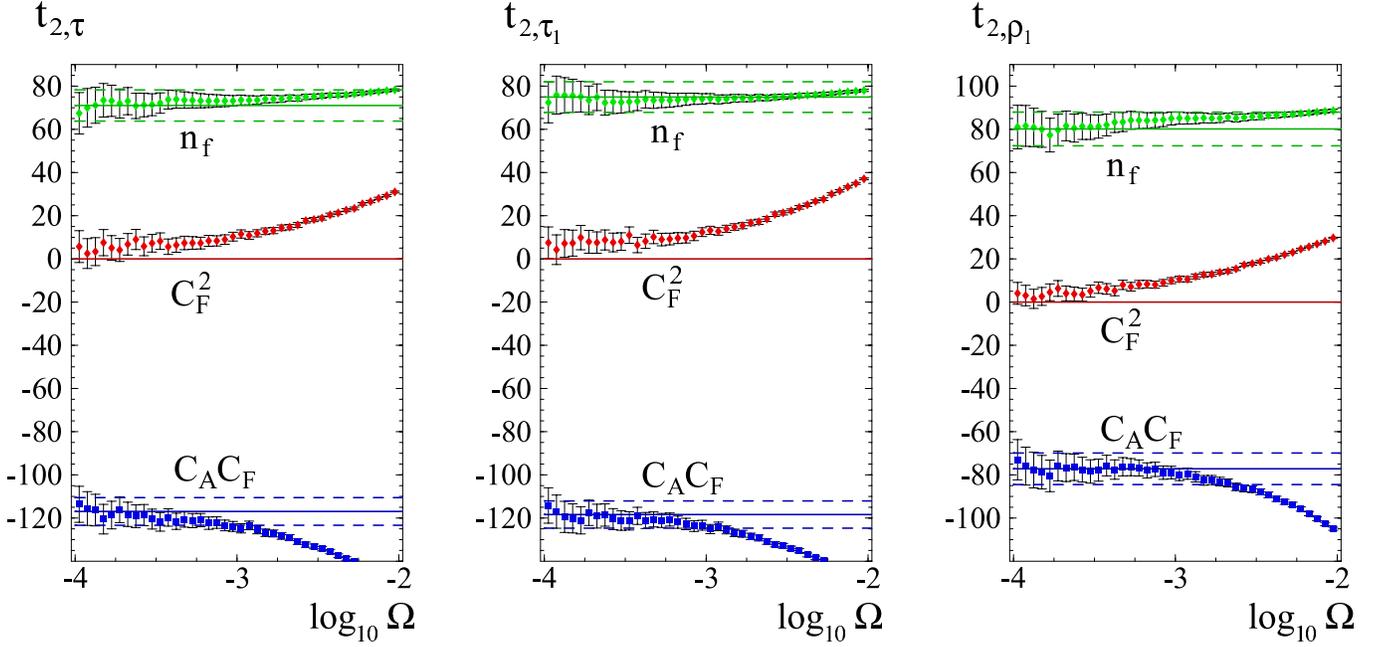}
\vskip -0.4cm
\caption{
Results (solid and dashed lines) and estimates for $t_{2,y}$,
$y=\tau, \tau_\alpha, \rho_\alpha$, for $n_f=5$    
(using a presentation analogous to the one in the right column of
Fig.~\ref{fig:aeq1}) separated according the color factor 
contributions proportional to $C_A C_F$ (blue), $C_F^2$ (red) and $C_F T n_f$
(green). In the limit $\Omega\to 0$ the $C_F^2$ contributions are zero due to
nonabelian exponentiation, and no error estimate is given.}
\label{fig:colorfactors}
\end{center}
\end{figure}

Figure~\ref{fig:colorfactors} shows the results from last column in
Fig.~\ref{fig:aeq1} for $t_{2,\tau},t_{2,\tau_\alpha}$ and $t_{2,\rho_\alpha}$
separated according to the three types of color factors.\footnote{Since we
found numerical instabilities for the $C_F T n_f$ contributions obtained from 
EVENT2 for $\log_{10}\Omega\lsim -3.25$
we determined these contributions from subtracting the $C_F^2$ and $C_A C_F$
terms from the results for the sum of all color factors.} It is conspicuous
that the different color factor components approach their asymptotic value at
$\Omega=0$ at much smaller values for $\Omega$ as compared to
the sum of all color factors shown in
Fig.~\ref{fig:aeq1}. In particular, for the $C_F^2$ contributions the
compatibility with zero becomes only apparent for $\log_{10} \Omega\approx -4$.
Thus our results for the color factor components have somewhat
larger errors than for their sum that was obtained above. To obtain our
results we  use the average of the lowest five bins with $\log_{10}\Omega\le -3.75$
and adopt the error at $\log_{10}\Omega=-3.75$ as the uncertainty. For the color
factor contributions of $s_1$ and $s_2$ we find
\begin{align}
\label{abcolorfactors}
& s_{1,C_F^2} \, = \, s_{2,C_F^2} \, = 0\,,
\nn\\[1mm]
&s_{1,C_A C_F} \,  = \, -117.6 \pm 4.5 
\,, &&
s_{2,C_A C_F} \, = \, -12.3 \pm 2.6 
\,,
\nn\\[1mm]
&s_{1,n_f} \,  = \, \left\{\begin{array}{lr} 73.0 \pm 5.1 & (n_f=5) \\ 
59.5 \pm 5.1 & (n_f=4)\end{array}\right.
\,, &&
s_{2,n_f} \, = \, \left\{\begin{array}{lr} -2.2 \pm 2.8 & (n_f=5) \\ 
-1.8 \pm 2.8 &  (n_f=4)\end{array}\right.
\,,
\end{align}
where we have adopted zero for the $C_F^2$ contributions as required by the
nonabelian exponentiation property. As for our
results for the sum of 
all color factors, we have checked that the results in
Eq.~(\ref{abcolorfactors}) are fully compatible with the color factor
contributions of $t_{2,{\rm \tau_\alpha}}$ and  $t_{2,{\rm \rho_\alpha}}$
obtained from EVENT2 for $\alpha=2,3,5,10$ and $1/2,1/3,1/5,1/10$. 
A determination of the color factor components of $s_1$ for was also
carried out 
in Ref.~\cite{Becher:2008cf}. Transferred to our notation their result reads  
$s_{1,C_F^2}=8.3 \pm 1.8$, $s_{1,C_A C_F}=-120.0 \pm 2.0$, 
$s_{1,n_f=5}=71.7 \pm 1.7$ and $s_{1,n_f=4}=57.3 \pm 1.3$.
For the $C_F^2$ term a quite small error is claimed, rendering the
result incompatible with zero. This is because in Ref.~\cite{Becher:2008cf}
EVENT2 was run with linear binning, and 
only results for $\log_{10}\Omega\ge -3$ were used for the analysis assuming that
the asymptotic values can be extrapolated from them. While our analysis
validates this approach for the sum of all color factors, it can be seen from   
Fig.~\ref{fig:colorfactors} that it fails for the $C_F^2$
color factor contribution of $s_1$. For the $C_AC_F$ and the $n_f$
contributions of $s_1$ the  results of Ref.~\cite{Becher:2008cf} are
compatible with ours, but we believe that our error estimate is more
appropriate.

\vskip 5mm
\section{Soft Function Gap}
\label{sectiongap}

The curves in Fig.~\ref{fig:softpure} show a rather poor behavior of the shape
of the soft function in perturbation theory. This behavior can be improved
considerably when subtractions are applied to the partonic soft function
$S_{\rm part}$ that remove the ${\cal O}(\Lambda_{\rm QCD})$ renormalon in the
partonic threshold at $\ell^\pm=0$~\cite{Hoang:2007vb}.\footnote{
This renormalon in the soft function is the origin of the ${\cal
  O}(\Lambda_{\rm QCD})$ renormalon identified in Ref.~\cite{Gardi:2001ny} 
in the perturbative expansion of the thrust distribution in full QCD.
} This
renormalon is very similar in nature (but independent of) the well-known
${\cal O}(\Lambda_{\rm QCD})$ pole mass renormalon in the heavy quark mass
threshold in the partonic on-shell limit
$q^2-m^2=0$~\cite{Bigi:1994em,Beneke:1994sw}.   
To remove this renormalon order-by-order in the perturbative expansion it was
suggested in Ref.~\cite{Hoang:2007vb} to introduce the scale-independent gap
parameter $\Delta$ in the soft function model function as shown in
Eq.~(\ref{Sgap}). At this level $\Delta$ represents an additional model
parameter that can compensate numerically for the divergent higher order
behavior of the soft function caused by the ${\cal O}(\Lambda_{\rm QCD})$
renormalon. To obtain a gap parameter that is more stable in perturbation
theory and allows a meaningful determination from experimental data it is
mandatory to remove the renormalon contributions in the partonic soft function
by explicit subtractions. This can be achieved by writing
\begin{align}
\Delta=\bar\Delta+\delta\bar\Delta\,,
\end{align}
where $\delta\bar\Delta$ is a perturbative series 
\begin{align}
\delta\bar\Delta \, = \, 
\delta\bar\Delta_1 \, + \,
\delta\bar\Delta_2 \, + \,
\delta\bar\Delta_3 \, + \,\ldots
\,,
\end{align}
that contains exactly the same ${\cal O}(\Lambda_{\rm QCD})$ renormalon as the
soft function. Starting from Eq.~(\ref{Sdef2}) and shifting the integration
variables the soft function can then be rewritten as
\begin{align}
\label{Sdef3}
S(\ell^+,\ell^-,\mu) \, = \,
\int_{-\infty}^{+\infty}\!\!\! d\bar\ell^+
\int_{-\infty}^{+\infty}\!\!\! d\bar\ell^-\
S_{\rm part}(\ell^+ \minus \bar\ell^+ \minus \delta\bar\Delta,
  \ell^- \minus \bar\ell^-\minus \delta\bar\Delta,\mu)\,
f_{\rm exp}(\bar\ell^+ \minus \bar\Delta,\bar\ell^- \minus \bar\Delta) \,.
\end{align}
To cancel the renormalon between the partonic soft function and
the series $\delta\bar\Delta$ order-by-order in the $\alpha_s$ expansion we
now have to expand Eq.~(\ref{Sdef3}) in the $\delta\bar\Delta_i$
simultaneously with the expansion for the partonic soft function $S_{\rm
part}=S_{\rm part}^0+S_{\rm   part}^1+S_{\rm   part}^2+\ldots$, so that
\begin{align}
\label{Sdef4}
 S_{\rm part}(\ell^\pm - \delta\bar\Delta,\mu)
& \,  = \,
S_{\rm part}^0(\ell^\pm,\mu) + \bigg[ S_{\rm part}^1(\ell^\pm,\mu) -\delta\bar\Delta_1 
   \Big(\frac{d}{d\ell^+}\plus \frac{d}{d\ell^-}\Big) S_{\rm part}^0(\ell^\pm,\mu)
   \bigg]\nn\\
 &\qquad + \bigg[ S_{\rm part}^2(\ell^\pm,\mu) -
   \Big(\frac{d}{d\ell^+}\plus \frac{d}{d\ell^-}\Big) 
\Big\{ \delta\bar\Delta_2 S_{\rm part}^0(\ell^\pm,\mu) 
+\delta\bar\Delta_1 S_{\rm part}^1(\ell^\pm,\mu) \Big\}
\nn\\[1mm]
 &\qquad\quad + 
   \Big(\frac{d^2}{d\ell^{+\,2}}\plus \frac{d^2}{d\ell^{-\,2}} \plus 
   2 \frac{d^2}{d\ell^+d\ell^-}\Big) \frac{\delta\bar\Delta_1^{\, 2}}{2} 
   S_{\rm part}^0(\ell^\pm,\mu)
   \bigg] \, + \,\ldots
\,,
\end{align}
where $\delta\bar\Delta_i$ and $S_{\rm part}^i$ are of ${\cal
O}(\alpha_s^i)$. We stress that it is mandatory to use the same
renormalization scale $\mu$ for the expansion of $\delta\bar\Delta$ and
$S_{\rm part}$ to achieve the renormalon cancellation, see
e.g.\,\,Ref.~\cite{Hoang:2005zw}.  In Ref.~\cite{Hoang:2007vb} a 
definition of the series in $\delta\bar\Delta$ was proposed based on a ratio
of moments of the partonic soft function with a finite cutoff. Since the soft
function has an anomalous dimension, $\delta\bar\Delta$ and therefore also
$\bar\Delta$ are $\mu$-dependent. The moment definition
does, however, not allow to formulate a consistent RG running of the gap
parameter $\bar\Delta$ due to logarithmic terms of arbitrary high powers in
the evolution equations at high orders and because the resulting evolution
equation is not transitive~\cite{Jain:2008gb}.   
\\

\noindent
{\it Position space gap parameter.} We can define a gap parameter with a
consistent RG evolution from the position space soft function
\begin{align}
\label{Sdef5}
S(x_1,& x_2,\mu) \, = 
\int\!\! d\ell^{+} d\ell^{-} \:
e^{-i \,\ell^+ x_1}\,e^{-i\,\ell^- x_2}\,
S(\ell^+,\ell^-,\mu)
\nn\\[1mm] &\, = \,
\int\!\! d\ell^{+} d\ell^{-} \:
e^{-i \,\ell^+ x_1}\,e^{-i\,\ell^- x_2}\,
\int_{-\infty}^{+\infty}\!\!\! d\ell^{\prime +}
\int_{-\infty}^{+\infty}\!\!\! d\ell^{\prime -}\
S_{\rm part}(\ell^+ \minus \ell^{\prime +}-\delta\bar\Delta,
\ell^- \minus \ell^{\prime -}-\delta\bar\Delta,\mu)\,
f_{\rm exp}(\ell^{\prime +}-\bar\Delta,\ell^{\prime -}-\bar\Delta) 
\nn\\[1mm] &\, = \,
\tilde S_{\rm part}(x_1,x_2,\mu)\,
\left[\, f_{\rm exp}(x_1,x_2)\,e^{-i\bar\Delta(x_1+x_2)} \,\right]\,,
\end{align}
where 
\begin{align}
f_{\rm exp}(x_1,x_2) \, = &\,\, 
\int\!\! d\ell^{+} d\ell^{-} \:
e^{-i \,\ell^+ x_1}\,e^{-i\,\ell^- x_2}\,
f_{\rm exp}(\ell^+,\ell^-)
\end{align}
and
\begin{align}
\tilde S_{\rm part}(x_1,x_2,\mu)\,= &\,\, 
S_{\rm part}(x_1,x_2,\mu)\, e^{-i \delta\bar\Delta(x_1+x_2)}
\,.
\end{align}
Since the function $\tilde S_{\rm part}(x_1,x_2,\mu)$ is supposed to be free
of the ${\cal O}(\Lambda_{\rm QCD})$ renormalon we can use the condition
\begin{align}
\label{gapdef1}
R\,e^{\gamma_E}\,\frac{d}{d\ln(i x_1)}\,
\left.\left[\,\ln \tilde S_{\rm part}(x_1,x_2,\mu)\,\right]\right|_{x_1=x_2=(i R
e^{\gamma_E})^{-1}}\, = \, 0
\end{align}
to derive an explicit expression for $\delta\bar\Delta$,
\begin{align}
\label{gapdef2}
\delta\bar\Delta(R,\mu)\,&\, = \,
R\,e^{\gamma_E}\,\frac{d}{d\ln(i x_1)}\,
\left.\Big[\,\ln S_{\rm part}(x_1,x_2,\mu)\,\Big]\right|_{x_1=x_2=(i R
e^{\gamma_E})^{-1}}
\nn\\[1mm] &
\, = \,
\delta\bar\Delta_1(R,\mu) \, + \,
\delta\bar\Delta_2(R,\mu) \, + \,
\delta\bar\Delta_3(R,\mu) \, + \, \ldots
\,,
\end{align}
where the scale $R$ is free parameter.
This position space method was used before in Ref.~\cite{Jain:2008gb} to
derive a short-distance jet mass definition from the jet functions $B_\pm$
that appear in the factorization theorem~(\ref{facttheobHQET}) for massive
quarks in the resonance region. The explicit results for the
$\delta\bar\Delta_i$'s up to ${\cal O}(\alpha_s^3)$ read 
[$L_{\mu R}\equiv\ln(\frac{\mu}{R})$, $\alpha_s=\alpha_s(\mu)$]
\begin{align}
\label{deltaDeltadef}
\delta &\bar\Delta_1(R,\mu) \, = \,
\left(\frac{\alpha_s}{4\pi}\right)\,
\Big[\,
\gamma_s^0 + 2 \Gamma_s^0 \,L_{\mu R}
\,\Big]
\, = \,
\left(\frac{\alpha_s}{4\pi}\right)\,
\Big[\!
-8\, C_F\,L_{\mu R}
\,\Big]\,,
\nn\\[1mm]
\delta &\bar\Delta_2(R,\mu) \, = \,
\left(\frac{\alpha_s}{4\pi}\right)^2\,
\Big[\,
2 t_1 \beta_0 + \gamma_s^1 
+ 2\Big(\beta_0 \gamma_0 + \Gamma_s^1\Big)L_{\mu R} 
+  2\beta_0\Gamma_s^0 \,L_{\mu R}^2
\,\Big]
\nn\\[1mm] &
\, = \,
\left(\frac{\alpha_s}{4\pi}\right)^2\,
\bigg[\,
C_A\,C_F\,\bigg(\!\! 
  -\frac{808}{27} - \frac{22}{9}\pi^2 + 28\zeta_3  
  + \Big(\!\!-\frac{536}{9} + \frac{8}{3}\pi^2\Big)L_{\mu R}- \frac{88}{3}L_{\mu R}^2 \bigg)
\nn\\[1mm] & \mbox{\hspace{1.5cm}} 
+ C_F\,T n_f\,\bigg( \frac{224}{27} + \frac{8}{9}\pi^2 
  + \frac{160}{9}\,L_{\mu R} + \frac{32}{3}\,L_{\mu R}^2 \bigg) 
\,\bigg]\,,
\nn\\[1mm]
\delta &\bar\Delta_3(R,\mu) \, = \,
\left(\frac{\alpha_s}{4\pi}\right)^3\,
\Big[\,
4 s_1\beta_0 + 2 t_1\beta_1 + \gamma_s^2 
+ 2 \Big(4 t_1\beta_0^2 + \beta_1\gamma_s^0 + 2\beta_0\gamma_s^1 +
\Gamma_s^2\Big)L_{\mu R}
\nn\\[1mm] & \mbox{\hspace{3.0cm}} 
+  2 \Big(2\beta_0^2\gamma_s^0 + \beta_1\Gamma_s^0 
        + 2\beta_0\Gamma_s^1\Big)L_{\mu R}^2 
+ \frac{8}{3}\beta_0^2\Gamma_s^0 \,L_{\mu R}^3
\,\Big]
\nn\\[1mm] &
\, = \,
\left(\frac{\alpha_s}{4\pi}\right)^3\,
\bigg[\,
 C_A^2\,C_F\,\bigg(\!\! - \frac{34}{3}\pi^2  
  + \bigg(\!\!-\frac{62012}{81} + \frac{104}{27}\pi^2 - \frac{88}{45}\pi^4 
           + 352\,\zeta_3\,\bigg) L_{\mu R}
  + \bigg(\!\!-\frac{14240}{27} + \frac{176}{9}\pi^2\bigg)L_{\mu R}
  -\frac{3872}{27}L_{\mu R}^3
  \bigg)
\nn\\[1mm] & \mbox{\hspace{1.5cm}} 
+ C_A\,C_F\,T n_f\,\bigg( \,\frac{20}{3}\pi^2 
    + \bigg(\,\frac{32816}{81} + \frac{128}{9}\pi^2\bigg) L_{\mu R}  
       + \bigg(\,\frac{9248}{27} - \frac{64}{9}\pi^2\bigg)L_{\mu R} 
       + \frac{2816}{27}L_{\mu R}^3 \bigg)
\nn\\[1mm] & \mbox{\hspace{1.5cm}} 
+ C_F^2\,T n_f\,\bigg( 4\pi^2 + \bigg(\frac{440}{3} - 128\zeta_3\bigg)L_{\mu R} 
        + 32\, L_{\mu R}^2  \bigg) 
\nn\\[1mm] & \mbox{\hspace{1.5cm}} 
+ C_F\,(T n_f)^2 \bigg( \bigg(\!\!-\frac{3200}{81} -
       \frac{128}{27}\pi^2\bigg)L_{\mu R} -\frac{1280}{27}L_{\mu R}^2 
     - \frac{512}{27}L_{\mu R}^3 
    \bigg)
+ 4\,s_1\,\bigg( \frac{11}{3}C_A-\frac{4}{3}T n_f\bigg)
+\gamma_s^2 
\,\bigg]
\,.
\end{align}
The ${\cal O}(\alpha_s^3)$ corrections is only partially known as it depends
on the unknown 3-loop non-cusp anomalous dimension $\gamma_s^2$ of the soft
function. Using that $\Delta=\bar\Delta+\delta\bar\Delta$ is scale-independent
it is straightforward to derive the anomalous dimension of the gap parameter
$\bar\Delta(R,\mu)$, which turns out to be just proportional to the cusp
anomalous dimension of the soft function in Eq.~(\ref{cuspnoncusp1}),
\begin{align}
\frac{d}{d\ln\mu}\,\bar\Delta(R,\mu) & \, = \,
-\frac{d}{d\ln\mu}\,\delta\bar\Delta(R,\mu)
\, = \,
-2\,R\,e^{\gamma_E}\,\Gamma_s[\alpha_s(\mu)]
\,.
\end{align}
Up to ${\cal O}(\alpha_s^3)$ this  intriguing fact can derived explicitly from
the results in Eqs.~(\ref{deltaDeltadef}). To all orders it can be proven
using the expression~(\ref{UUS3}) for the position space soft function and the
all-orders definition for $\tilde K$ in Eq.~(\ref{omegaK}).
The solution of the RG equation reads
\begin{align}
\label{Deltabarmusolved}
\bar\Delta(R,\mu) \, = \, \bar\Delta(R,\mu_0) -
R\,e^{\gamma_E}\,\tilde\omega(\Gamma_s,\mu,\mu_0) 
\,,
\end{align}
where $\tilde\omega$ is given in Eq.~(\ref{omegaK}).
\\

\noindent
{\it R-Evolution} 
The definition of the gap in Eq.~(\ref{gapdef2}) depends on the
choice of the scale parameter $R$. Since $\delta\bar\Delta(R,\mu)$ represents
an infrared subtraction of low-energy fluctuations from the partonic soft
function, one can interpret the scale $R$ as an infrared cutoff governing
infrared fluctuations that are absorbed into the renormalon-free gap parameter
$\bar\Delta(R,\mu)$. In this respect $R$ differs significantly from the
renormalization scale $\mu$, which governs ultraviolet fluctuations that are
absorbed into coupling constants. Like the renormalization scale $\mu$, the
scale $R$ 
should be taken in the perturbative regime, to allow for a perturbative
description of the evolution of $\bar\Delta(R,\mu)$ in $R$, but it should also
be close to the typical scales that are governing the soft function. 
Moreover to avoid large logs $\mu/R$ should be of order one. For the
description of the peak region for massless jet event shapes or the massive
jet invariant mass resonance region one is sensitive to details of the shape
of the soft function.  So the typical range of scales for $\mu$ and $R$ in
this case is around $1-1.5$~GeV. For describing these distributions in
the tail away from the peak and resonance region one can use an operator
product expansion and only global properties of the soft function in the form
of moments are relevant. So here the perturbative contributions are dominated
by larger scales, and for $\mu$ and $R$ larger scales should be adopted as
well.\footnote{
Because the one-loop subtraction $\delta\bar\Delta_1$ term only contains a
logarithm of $\mu/R$ it is also required in practice to choose  $R<\mu$
to have a subtraction with the correct sign at the one-loop level.
} 

Besides these constrains the choices of $\mu$ and $R$ are arbitrary. It is
therefore useful to also have an evolution equation for the gap parameter with
respect to $R$. A detailed study of the required formalism for evolving
$\bar\Delta(R,R)$ in $R$ and its relation to renormalons was carried out
recently in Ref.~\cite{Hoang:2008yj} and we refer the interested reader to
this work. For the renormalon-free soft function gap parameter the 
$R$-evolution equation has the form
\begin{align}
\label{DeltbarRRGE}
\frac{d}{d R}\,\bar\Delta(R,R) \, = \, 
-\,R\,\sum_{n=0}^{\infty} 
\left(\frac{\alpha_s(R)}{4\pi}\right)^{n+1}\,\gamma_n^R
\, = \, 
-\,R\,\bigg[\, 
\left(\frac{\alpha_s(R)}{4\pi}\right)\,\gamma_0^R \, + \,
\left(\frac{\alpha_s(R)}{4\pi}\right)^{2}\,\gamma_1^R \, + \,
\left(\frac{\alpha_s(R)}{4\pi}\right)^{3}\,\gamma_2^R \, + \,
\ldots\
\,\bigg]
\,,
\end{align}
where the first three terms can be obtained from Eq.~(\ref{deltaDeltadef}) and
read 
\begin{align}
\label{gammaR}
\gamma_0^R & \, = \, 0
\,,
\nn\\[1mm]
\gamma_1^R & \, = \,
C_A\,C_F\,\bigg(\!\!-\frac{808}{27} - \frac{22}{9}\pi^2 + 28\zeta_3\bigg)
\, + \,
C_F\,T n_f \bigg(\frac{224}{27} + \frac{8}{9}\pi^2\bigg)
\,,
\nn\\[1mm]
\gamma_2^R & \, = \,
C_A^2\,C_F\,\bigg(\frac{35552}{81} + \frac{662}{27}\pi^2 -
      \frac{1232}{3}\zeta_3\bigg) 
+ C_A\,C_F\,T n_f\,\bigg(\!\!-\frac{22784}{81} - \frac{524}{27}\pi^2 
      + \frac{448}{3}\zeta_3\bigg)
\nn\\[1mm] & \mbox{\hspace{0.5cm}}
+ C_F\,(T n_f)^2\,\bigg(\frac{3584}{81} + \frac{128}{27}\pi^2\bigg)
+ 4\,C_F^2\,T n_f\,\pi^2 
+ 4s_1\,\bigg(\frac{11}{3} C_A - \frac{4}{3}T n_f\bigg)
+ \gamma_s^2
\,. 
\end{align}
At N${}^k$LL order the analytic solution reads 
[$t_R\equiv -2\pi/(\beta_0\alpha_s(R))$] 
\begin{align}
\label{DeltabarRsolved}
 & \big[\bar\Delta(R,R)\minus \bar\Delta(R_0,R_0)\big]^{{\rm N}^k{\rm LL}}
  = \Lambda^{(k)}_{\rm QCD}\,\sum_{j=0}^k \,\  S_j\, (-1)^j 
  \, e^{i\pi \hat b_1} \, \Big[ \, \Gamma(-\hat b_1\minus j,t_R)
  - \Gamma(-\hat b_1 \minus j, t_{R_0}) \,\Big]
  \,,
\end{align}
where [$\tilde\gamma_n^R\equiv \gamma_n^R/(2\beta_0)^{n+1}$]
\begin{align}
 S_0 &\,=\, \tilde \gamma_0^R \,, 
 \nn\\[1mm]
 S_1 &\,=\, \tilde \gamma_1^R - \Big(\hat b_1\plus \hat b_2\Big) \tilde \gamma_0^R\,,
 \nn\\[1mm]
 S_2 &\,=\, \tilde \gamma_2^R 
     - \Big(\hat b_1\plus \hat b_2\Big) \tilde \gamma_1^R
    +\left[ \Big(1\plus \hat b_1\Big)\hat b_2 +\frac{1}{2}\left(\hat b_2^2 \plus \hat b_3\right) \right]
  \tilde \gamma_0^R .
\end{align}
The terms $\hat b_i$ are the coefficient of the large-$t$ expansion of the
function $\hat b(t)=1+\hat b_1/t + \hat b_2/t^2 + \ldots$ that appears in the relation
\begin{align}
 \ln\frac{R}{R_0} =\!\! \int_{\alpha_s(R_0)}^{\alpha_s(R)} \frac{d\alpha_R}{\beta[\alpha_R]} 
  =\! \int_{t_R}^{t_{R_0}}\!\!\! dt\: \hat b(t) = G(t_{R_0}) - G(t_R) \,,
\end{align}
where the first few terms read
\begin{align}
\hat b_1 \, =\, \frac{\beta_1}{2\beta_0^2}\,,
\mbox{\hspace{1.5cm}}
\hat b_2 \, = \, \frac{\beta_1^2-\beta_0\beta_2}{4\beta_0^4}\,,
\mbox{\hspace{1.5cm}} 
\hat b_3 \, = \,
\frac{\beta_1^3-2\beta_0\beta_1\beta_2+\beta_0^2\beta_3}{8\beta_0^6}
\,,
\end{align}
and $\Lambda_{\rm QCD}^{(k)}$ is the N${}^k$LL order approximation of 
\begin{align}
\Lambda_{\rm QCD} \, = \, R\,e^{G(t_R)} \, = \, R_0\,e^{G(t_{R_0})}
\end{align}
which is the familiar definition of $\Lambda_{\rm QCD}$ from the strong
coupling. 

From the solutions for $\mu$- and $R$-evolution given in
Eqs.~(\ref{Deltabarmusolved}) and (\ref{DeltabarRsolved})  one can relate
$\bar\Delta$ for arbitrary $R$ and $\mu$ values through the relation
\begin{align}
\label{Deltabarfullysolved}
\bar\Delta(R,\mu) \, = \,&\, \bar\Delta(R_0,\mu_0) - R\,e^{\gamma_E}\,\Big[\,
\tilde\omega(\Gamma_s,R_0,\mu_0) \, + \, \tilde\omega(\Gamma_s,\mu,R)
\,\Big] 
\nn\\[1mm] &
\, + \,
\Lambda_{\rm QCD}\,\sum_{j=0}^\infty\  S_j\, (-1)^j 
  \, e^{i\pi \hat b_1}  \Big[ \Gamma(-\hat b_1\minus j,t_R)
  - \Gamma(-\hat b_1 \minus j, t_{R_0}) \Big]
\end{align}

\vskip 5mm
\section{Numerical Results}
\label{sectionnumerical}

A detailed analysis of the impact of the renormalon subtraction for the soft
function has been given recently in Ref.~\cite{Hoang:2007vb} using an
approximation for the at that time unknown ${\cal O}(\alpha_s^2)$
corrections. Since for the full ${\cal O}(\alpha_s^2)$ soft function
and the gap parameter obtained in 
this work we find qualitatively similar results, we do not repeat such an
extended analysis here. In the following brief numerical analysis we intend to 
illustrate the impact for a few examples and to demonstrate the importance of
accounting for the evolution of the gap parameter $\bar\Delta(R,\mu)$ in
$(R,\mu)$-space, which was not available for the gap parameter used in
Ref.~\cite{Hoang:2007vb}. 

\begin{figure}[t]
\begin{center}
\epsfxsize=\textwidth
\epsffile{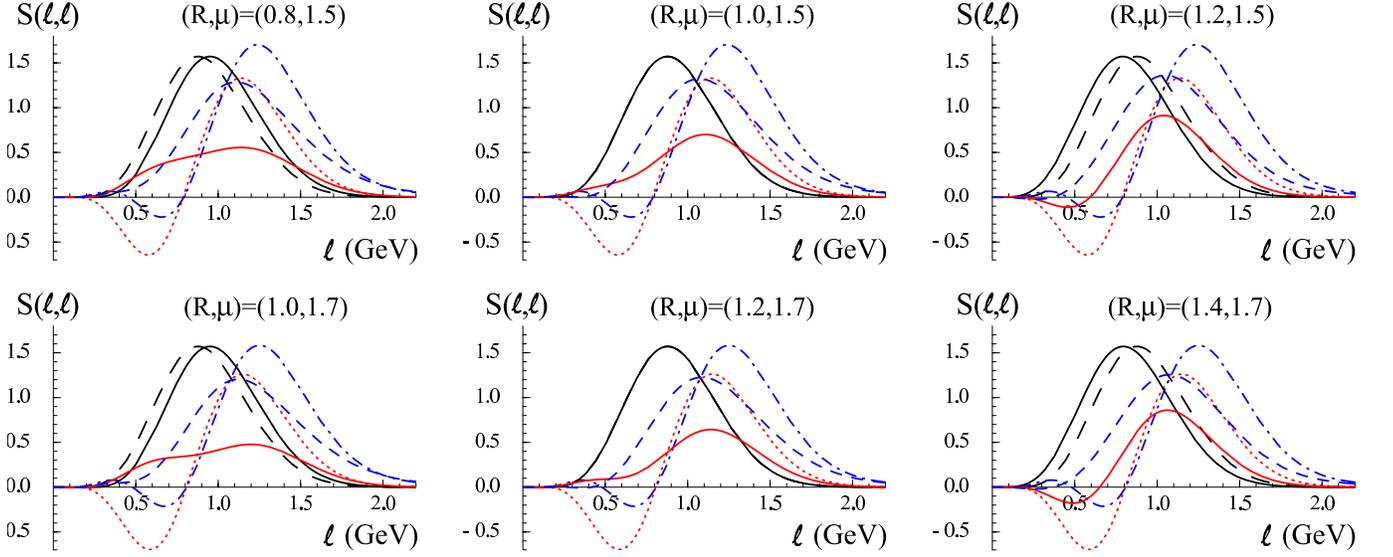}
\vskip -0.3cm
\caption{
Dependence of the soft function $S(\ell,\ell,\mu)$ on the
renormalization scale $\mu$ and the infrared subtraction scale $R$ 
at tree-level, one-loop and two-loop with and without renormalon subtraction.  
In the middle column the solid and long-dashed black lines coincide.
See the text for details. 
 }
\label{fig:softscan}
\end{center}
\end{figure}

In Fig.~\ref{fig:softscan} the soft function $S(\ell,\ell,\mu)$ is
plotted with and without renormalon subtraction for $\mu=1.5$~GeV and
$1.7$~GeV ($\alpha_s(1.5~\mbox{GeV})=0.3285$) and different choices of $R$ for
the renormalon subtracted soft function. For all curves we have used $n_f=5$
and the soft function model $S_{\rm mod}$ with the parameters
$\Lambda=0.55$~GeV and $(a,b)=(3.0,-0.5)$. For the soft 
function without renormalon subtraction (i.e.\,\,$\delta\bar\Delta=0$ and
$\Delta=\bar\Delta$) we have used $\Delta=0.1$~GeV, and the tree-level
(long-dashed black lines), ${\cal O}(\alpha_s)$ (dotted red lines) and ${\cal
O}(\alpha_s^2)$ (dotted-dashed blue lines) results are displayed. The results
show the rather poor perturbative behavior already observed before in
Fig.~\ref{fig:softpure}. In particular, the soft functions at ${\cal O}(\alpha_s)$
and ${\cal O}(\alpha_s^2)$ have unphysical negative values for small values of
$\ell$. For the soft function with renormalon subtraction we have used
$\bar\Delta(1.0~\mbox{GeV},1.5~\mbox{GeV})=0.1$~GeV and computed $\bar\Delta$
for the $(R,\mu)$ values in the different panels using
Eq.~(\ref{Deltabarfullysolved}) at NNLL order.\footnote{
We have set the unknown non-cusp term $\gamma_s^2$ to zero for the NNLL order
R-evolution of $\bar\Delta(R,\mu)$.} 
In the different panels the tree-level (solid black lines), ${\cal
O}(\alpha_s)$ (lighter solid red lines) and ${\cal O}(\alpha_s^2)$ (dashed
blue lines) results are displayed. The results for the soft function with
renormalon subtractions show a substantially improved perturbative behavior.
Concerning the shape and the peak location, the ${\cal
O}(\alpha_s^2)$ corrections are particularly small. It is also conspicuous 
that the curves do not have unphysical negative values except for
$(R,\mu)=(1.5,1.2)$~GeV and $(1.7,1.4)$~GeV where the difference of $R$ and
$\mu$ is small so that the subtractions at ${\cal O}(\alpha_s)$ are 
not sufficiently large (see footnote~1). 

\begin{figure}[t]
\begin{center}
\epsfxsize=10cm
\epsffile{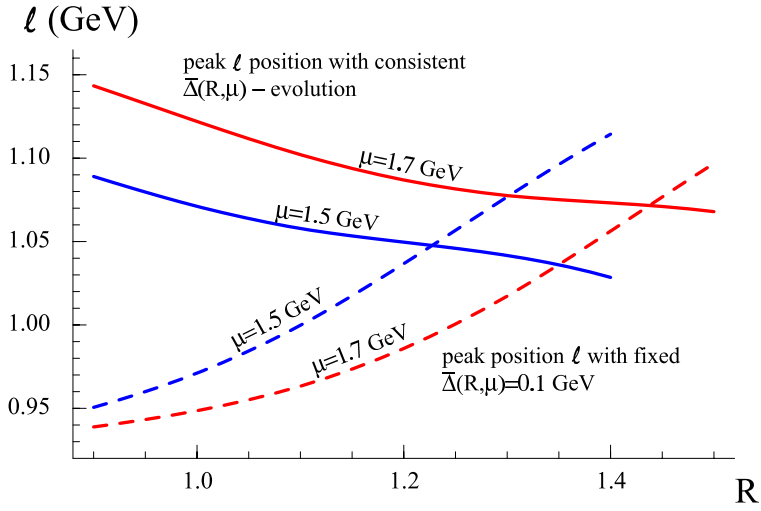}
\vskip -0.3cm
\caption{Location of the maximum of the soft function $S(\ell,\ell,\mu)$ with
  renormalon 
  subtraction as a function of $R$ for $\mu=1.5$ and $1.7$~GeV. The solid
  curves account for the correct evolution of the gap parameter
  $\bar\Delta(R,\mu)$, while for the dashed curves a fixed gap parameter
  $\bar\Delta=0.1$~GeV is used for all values of $R$ and $\mu$.}
\label{fig:peakpos}
\end{center}
\end{figure}

For the stabilization of the peak position with respect to variations of $R$
it is important to use the consistent evolution of $\bar\Delta$ in the
$(R,\mu)$-plane. In Fig.~\ref{fig:peakpos} the position of the maximum of the
${\cal O}(\alpha_s^2)$ soft function with renormalon subtractions is displayed
as a function of $R$ for the soft function model used also in
Fig.~\ref{fig:softscan} and for $\mu=1.5$~GeV and $\mu=1.7$~GeV. The solid
lines show the peak location with a consistent choice of $\bar\Delta(R,\mu)$
using $\bar\Delta(1.0~\mbox{GeV},1.5~\mbox{GeV})=0.1$ and
Eq.~(\ref{Deltabarfullysolved}) to obtain $\bar\Delta$ for other
$(R,\mu)$-values. The dashed lines, on the other hand, show the results when
$\bar\Delta=0.1$~GeV for any $R$ and $\mu$. The curves show a considerably
weaker $R$-dependence when the correct evolution of the gap parameter
$\bar\Delta(R,\mu)$ is accounted for. Although this is expected due to the
fact  that $\Delta=\bar\Delta+\delta\bar\Delta$ is $R$-independent, it is
reassuring that this is also reflected in the final result for the soft
function given that $\delta\bar\Delta$ is implemented by subtractions that
change the shape of the soft function substantially, whereas $\bar\Delta$ is
implemented by a simple shift of variable $\ell$. Note that
$\Delta=\bar\Delta+\delta\bar\Delta$ is also $\mu$-independent. However, since
the soft function has a non-zero anomalous dimension, it is not expected that
accounting for the proper $\mu$-evolution of $\bar\Delta$ also leads to a
smaller $\mu$-dependence of the peak location. This is confirmed by the
vertical separation of the two solid curves that is in general not smaller
than for the corresponding two dashed curves. However, it is clearly visible
that the vertical separation for the solid curves (which account for the
consistent evolution of $\bar\Delta$) is approximately $R$-independent as
compared to the vertical separation of the dashed curves (which have a fixed
value of $\bar\Delta$). This shows that the variation of the location where
the soft function has its maximum with changes of $\mu$ is only
$R$-independent when the proper evolution of $\bar\Delta$ is included. 
This emphasizes once more the importance of having a consistent formulation of
the $R$- and $\mu$-evolutions of the gap parameter $\bar\Delta(R,\mu)$.

\vskip 5mm
\section{Conclusions}
\label{sectionconclusions}

In this work we have determined the partonic hemisphere soft
function $S_{\rm part}(\ell^+,\ell^-,\mu)$ at ${\cal O}(\alpha_s^2)$. The
partonic soft function $S_{\rm part}$ is an essential ingredient for the
theoretical description at NNLL order of thrust and heavy jet mass
distributions in the dijet and resonance limit, where perturbative and
nonperturbative contribution  can be separated according to the factorization
theorems~(\ref{facttheoSCET}) and (\ref{facttheobHQET}). Using general
properties of the partonic soft function originating from its renormalization
group structure and its exponentiation properties, we have been able to determine
the ${\cal O}(\alpha_s^2)$ corrections up to two constants which were
determined numerically using information from the MC program EVENT2 by Catani
and Seymour. 

To remove the ${\cal O}(\Lambda_{\rm QCD})$ renormalon of the partonic
threshold in  $S_{\rm part}$ at $\ell^\pm=0$ one can implement a subtraction
procedure devised in Ref.~\cite{Hoang:2007vb} that is based on a gap
parameter. Using this gap subtraction scheme one can remove order-by-order
this renormalon contribution in the perturbative parts of the factorization
theorem that otherwise causes instabilities for numerical determinations of
the first power correction from numerical data. 
From the result for the partonic soft function in position space 
representation we have defined such a gap subtraction scheme which has the
virtue of having a consistent renormalization group evolution in the
renormalization scale $\mu$ and the subtraction scale $R$. This property is
important for having a coherent theoretical description of event shape
distributions in the peak and the tail region.

\vskip 5mm
\section{Acknowledgements}
\label{sectionacknowledgements}

The authors would like to thank S.~Catani and M.~Seymour 
for helpful discussions on EVENT2. 
This work was supported in part by
the EU network contract MRTN-CT-2006-035482 (FLAVIAnet). 

\vskip 5mm

\appendix

\section{Hard Coefficient, Jet Function and Consistency Condition}
\label{app:hardjetcc}

\noindent
{\it Hard Coefficient $H_Q$.} 
The ${\cal O}(\alpha_s^2)$ fixed-order expression of the hard coefficient in
the dijet factorization theorem~(\ref{facttheoSCET}) for massless jets can be
derived from the 
results in Ref.~\cite{Matsuura:1988sm,Moch:2005id,Gehrmann:2005pd} and has the
form [$\alpha_s=\alpha_s(\mu)$, $L_Q\equiv \ln(Q^2/\mu^2)$]
\begin{align}
H_Q & (Q, \mu) \, = \, 1 
+ \left(\frac{C_F\alpha_s}{4\pi}\right) \,
  \bigg(\!\!- 2 L_Q^2 + 6 L_Q - 16 + \frac{7}{3}\pi^2\bigg)
\nn\\[1mm] &
+\,\left(\frac{\alpha_s}{4\pi}\right)^2 \,
\bigg\{
C_F^2\,\bigg[
  2 L_Q^4 - 12 L_Q^3 + \Big(50 - \frac{14}{3} \pi^2\Big) L_Q^2 
  + \Big(-93 + 10 \pi^2 + 48 \zeta_3\Big) L_Q
  + \frac{511}{4} - \frac{83}{3} \pi^2 + \frac{67}{30} \pi^4 - 60 \zeta_3
  \bigg]
\nn\\[1mm]&  \mbox{\hspace{0.8cm}}
+C_A\,C_F\,\bigg[ 
  \frac{22}{9} L_Q^3 + \Big(-\frac{233}{9} + \frac{2}{3} \pi^2\Big) L_Q^2 
  + \Big(\frac{2545}{27} - \frac{44}{9} \pi^2 - 52 \zeta_3\Big) L_Q 
  - \frac{51157}{324} + \frac{1061}{54} \pi^2 - \frac{8}{45} \pi^4 + \frac{626}{9} \zeta_3 
   \bigg]
\nn\\[1mm]&  \mbox{\hspace{.8cm}}
+C_F\,T\, n_f\,\bigg[\! 
   - \frac{8}{9} L_Q^3 + \frac{76}{9} L_Q^2 + \Big(-\frac{836}{27} 
   + \frac{16}{9} \pi^2 \Big) L_Q 
   + \frac{4085}{81} - \frac{182}{27} \pi^2 + \frac{8}{9} \zeta_3
   \bigg]
\bigg\}
\,.
\end{align}
The RG evolution is described by
$H_Q(Q,\mu)=H_Q(Q,\mu_0)U_{H_Q}(Q,\mu_0,\mu)$, where
\begin{align}
U_{H_Q}(Q,\mu_0,\mu) \, = \,& \,
\exp\bigg[
\tilde\omega(\Gamma_c,\mu,\mu_0)\ln\Big(\frac{\mu_0^2}{Q^2}\Big) 
\, +\,  2\tilde K(\Gamma_c,\gamma_c,\mu,\mu_0)
\bigg]\,.
\end{align}
The functions $\tilde\omega$ and $\tilde K$ are given in Eqs.~(\ref{omegaK})
and we have
\begin{align}
\Gamma_c^i \, = & \, - \Gamma_{\rm cusp}^i\,,\quad (i=0,1,\ldots)\,,
\nn \\[1mm]
\gamma_c^0 \, = & \,- 6 \,C_F
\,, 
\nn \\[1mm]
\gamma_c^1 \, = & \, \,C_F^2\Big(\!-3 + 4\pi^2 - 48\zeta_3\Big) 
  + C_A\,C_F \left(\!-\frac{961}{27} - \frac{11}{3}\pi^2 + 52 \zeta_3\right) 
  + C_F\, T n_f \left(\frac{260}{27} + \frac{4}{3}\pi^2\right)
\,,
\end{align}
where the terms in the cusp anomalous dimension $\Gamma_c$ are given in
Eq.~(\ref{Gammagamma}).
\\

\noindent
{\it Jet Function $J$.}
The ${\cal O}(\alpha_s^2)$ fixed-order expression for the  jet function
for massless quarks was computed in Ref.~\cite{Becher:2006qw}, where the
result was given in Laplace space. In momentum space the jet function reads\\[1mm]
[$\alpha_s=\alpha_s(\mu)$, 
${\cal L}^n_{s,+}=1/\mu^2[\theta(s)\ln^n(s/\mu^2)/(s/\mu^2)]_+$]
\begin{align}
J & (s,\mu)\, = \,\delta(s) 
+ \left(\frac{C_F\alpha_s}{4\pi}\right) \, 
\bigg( 4 {\cal L}^1_{s,+} - 3{\cal L}^0_{s,+} + (7-\pi^2)\,\delta(s)
\bigg)
\\[1mm] &
+\,\left(\frac{\alpha_s}{4\pi}\right)^2 \,
\bigg\{
C_F^2\,\bigg[
  8 {\cal L}^3_{s,+} - 18  {\cal L}^2_{s,+} 
  +\Big(37 - \frac{20}{3}\pi^2\Big) {\cal L}^1_{s,+}
  +\Big(-\frac{45}{2} + 7\pi^2 - 8\zeta_3\Big)  {\cal L}^0_{s,+}
\nn\\[1mm] & \mbox{\hspace{3.5cm}}
  +\Big(\frac{205}{8} - \frac{67}{6}\pi^2 + \frac{14}{15}\pi^4 - 18 \zeta_3\Big)\,\delta(s)
  \bigg]
\nn\\[1mm]&  \mbox{\hspace{0.8cm}}
+C_A\,C_F\,\bigg[ \!
 - \frac{22}{3} {\cal L}^2_{s,+} 
 + \Big(\frac{367}{9} - \frac{4}{3}\pi^2\Big) {\cal L}^1_{s,+}   
 + \Big(-\frac{3155}{54} + \frac{22}{9}\pi^2 + 40 \zeta_3\Big) {\cal L}^0_{s,+}
\nn\\[1mm] & \mbox{\hspace{3.5cm}}
 + \Big(\frac{53129}{648} - \frac{208}{27}\pi^2 - \frac{17}{180}\pi^4 
    - \frac{206}{9}\zeta_3\Big) \,\delta(s)
   \bigg]
\nn\\[1mm]&  \mbox{\hspace{.8cm}}
+C_F\,T\, n_f\,\bigg[
   \frac{8}{3} {\cal L}^2_{s,+}
 - \frac{116}{9} {\cal L}^1_{s,+}
 +  \Big(\frac{494}{27} - \frac{8}{9}\pi^2\Big) {\cal L}^0_{s,+}
 + \Big(-\frac{4057}{162} + \frac{68}{27}\pi^2 + \frac{16}{9}\zeta_3\Big)\,\delta(s)
   \bigg]
\bigg\}
\,,\nn
\end{align}
and its renormalization group evolution reads
$J(s,\mu) = \int ds^\prime U_J(s-s^\prime,\mu,\mu_0)J(s^\prime,\mu_0)$ with
\begin{align}
U_J(s,\mu,\mu_0) \, = \,
\frac{e^{\tilde K(\Gamma_J,\gamma_J,\mu,\mu_0)}
  (e^{\gamma_E})^{\frac{1}{2}\tilde\omega(\Gamma_J,\mu,\mu_0)}}
     {\mu_0^2\,\Gamma(-\frac{1}{2}\tilde\omega(\Gamma_J,\mu,\mu_0))}\,
 \bigg[ 
\frac{(\mu_0^2)^{1+\frac{1}{2}\tilde\omega(\Gamma_J,\mu,\mu_0)}\,\theta(s)}
{s^{1+\frac{1}{2}\tilde\omega(\Gamma_J,\mu,\mu_0)}}
 \bigg]_+
\,,
\end{align}
where the functions $\tilde\omega$ and $\tilde K$ are defined in
Eqs.~(\ref{omegaK}), the cusp and non-cusp anomalous dimensions are
given in Eqs.~(\ref{GammagammaJ}) and the definition of the plus function
given in Eq.~(\ref{plusfctdef}) is used.
In position space the jet function has the form [$L_\xi=\ln(i \xi e^{\gamma_E}\mu^2)$]
\begin{align}
J & (\xi, \mu) \, = \, 1 
+ \left(\frac{C_F\alpha_s}{4\pi}\right) \,
  \bigg( 2 L^2_\xi + 3 L_\xi + 7 - \frac{2}{3}\pi^2 \bigg)
\nn\\[1mm] &
+\,\left(\frac{\alpha_s}{4\pi}\right)^2 \,
\bigg\{
C_F^2\,\bigg[
  2 L_\xi^4 + 6L_\xi^3 + \Big(\frac{37}{2} - \frac{4}{3}\pi^2\Big) L_\xi^2 + 
  \Big(\frac{45}{2} - 4\pi^2 + 24\zeta_3\Big) L_\xi
  + \frac{205}{8}   - \frac{97}{12}\pi^2 + \frac{61}{90}\pi^4  - 6\zeta_3
  \bigg]
\nn\\[1mm]&  \mbox{\hspace{0.8cm}}
+C_A\,C_F\,\bigg[ 
 + \frac{22}{9} L_\xi^3 + \Big(\frac{367}{18} - \frac{2}{3} \pi^2\Big) L_\xi^2 
 + \Big(\frac{3155}{54} - \frac{11}{9} \pi^2 - 40 \zeta_3\Big) L_\xi
 + \frac{53129}{648}  - \frac{155}{36} \pi^2 - \frac{37}{180}\pi^4  + 
  - 18 \zeta_3
   \bigg]
\nn\\[1mm]&  \mbox{\hspace{.8cm}}
+C_F\,T\, n_f\,\bigg[\! 
 - \frac{8}{9} L_\xi^3 - \frac{58}{9} L_\xi^2 
 + \Big(-\frac{494}{27} + \frac{4}{9}\pi^2\Big) L_\xi
 - \frac{4057}{162}   + \frac{13}{9}\pi^2 
   \bigg]
\bigg\}
\,,
\end{align}
and its renormalization group evolution reads 
$J(\xi,\mu) = U_J(\xi,\mu,\mu_0)J(\xi,\mu_0)$ with
\begin{align}
U_J(\xi,\mu,\mu_0) \, = \, & \,
\exp\bigg[
\frac{1}{2}\tilde\omega(\Gamma_J,\mu,\mu_0)\ln\Big(i\xi e^{\gamma_E}\mu_0^2\Big) 
\, +\,  \tilde K(\Gamma_J,\gamma_J,\mu,\mu_0)
\bigg]\,.
\,,
\end{align}
where the functions $\tilde\omega$ and  
$\tilde K$ are given in Eqs.~(\ref{omegaK}). The cusp and non-cusp anomalous
dimensions read
\begin{align}
\label{GammagammaJ}
\Gamma_J^i \, = & \, 2 \Gamma_{\rm cusp}^i\,,\quad(i=0,1,\ldots)\,,
\nn \\[1mm]
\gamma_J^0 \, = & \,6 \,C_F
\,, 
\nn \\[1mm]
\gamma_J^1 \, = & \, \,C_F^2\Big( 3 - 4\pi^2 + 48\zeta_3\Big) 
  + C_A\,C_F \left(\frac{1769}{27} + \frac{22}{9}\pi^2 - 80 \zeta_3\right) 
  + C_F\, T n_f \left(\!\!-\frac{484}{27} - \frac{8}{9}\pi^2\right)
\,,
\end{align}
where the terms in the cusp anomalous dimension can be read of from
Eq.~(\ref{Gammagamma}). 

\vskip 2mm
\noindent
{\it Consistency Condition and Soft Function Evolution Factor $U_s$.} 
In Ref.~\cite{Fleming:2007xt} it was shown that the renormalization group
evolution $U$-functions of the hard coefficient $H_Q$, the jet function $J$
and the soft function $S$ in the factorization
theorem~(\ref{facttheoSCET}) for jets initiated by massless quarks are related
by a consistency condition. It can be
used to determine the renormalization group evolution functions of the soft
function. In momentum space the consistency condition reads
\begin{align}
U_s(\ell^\pm,\mu,\mu_0) \, = \, 
Q\,\sqrt{U_{H_Q}(Q,\mu,\mu_0)}\,U_J(Q\ell^\pm,\mu_0,\mu)
\,,
\end{align}
and in position space it has the form
\begin{align}
U_s(x_{1,2},\mu,\mu_0) \, = \, 
\sqrt{U_{H_Q}(Q,\mu,\mu_0)}\,U_J\Big(\frac{x_{1,2}}{Q},\mu_0,\mu\Big)
\,.
\end{align}
Using the relations
\begin{align}
\label{omegaKrelations}
\tilde\omega(\Gamma+\Gamma^\prime,\mu,\mu_0) \, = \,& \,
\tilde\omega(\Gamma,\mu,\mu_0) 
+ \tilde\omega(\Gamma^\prime,\mu,\mu_0)
\,,
\\[1mm]
\tilde K(\Gamma+\Gamma^\prime,\gamma+\gamma^\prime,\mu,\mu_0) \, = \,& \,
\tilde K(\Gamma,\gamma,\mu,\mu_0) 
+ \tilde K(\Gamma^\prime,\gamma^\prime,\mu,\mu_0)
\,,
\nn\\[1mm]
\tilde\omega(\Gamma,\mu,\mu^\prime) \, = \, & \,
-\,\tilde\omega(\Gamma,\mu^\prime,\mu)
\,,
\nn\\[1mm]
\tilde K(-\Gamma,\gamma,\mu_0,\mu)
-\tilde\omega\Big(\frac{\gamma}{2},\mu_0,\mu\Big) 
\, = \, & \,
-\tilde K(-\Gamma,\gamma,\mu,\mu_0)
+\tilde\omega\Big(\frac{\gamma}{2},\mu,\mu_0\Big)
-\tilde\omega(\Gamma,\mu,\mu_0)\ln\Big(\frac{\mu}{\mu_0}\Big)
\,,
\nn
\end{align}
it is then straightforward to derive Eqs.~(\ref{Usdefx}) and
(\ref{Usdefmom}). 
\\

\noindent
{\it Hard Coefficient $H_m$.} 
Recently the jet function $B_{\pm}$ was determined at ${\cal O}(\alpha_s^2)$
in Ref.~\cite{Jain:2008gb}. This jet function describes low-energy invariant mass
fluctuations of jets initiated by massive quarks in the factorization
theorem~(\ref{facttheobHQET}). Using the result for the NNLL anomalous dimension
for $B_\pm$ and the soft function $S$, and the consistency condition for the
$\mu$-evolution of the contributions in the factorization theorem for massive
jets in Eq.~(\ref{facttheobHQET}), one can derive the NNLL anomalous dimension
of the hard coefficient $H_m$ in Eq.~(\ref{facttheobHQET}). For the results
for the ${\cal O}(\alpha_s^2)$ jet function $B_\pm$ in momentum and position
space, which match exactly the conventions used here, we refer to
Ref.~\cite{Jain:2008gb}. The computation for the evolution factor for $H_m$ is
most conveniently done in position 
space representation. We start from 
the relation between the position and momentum space jet function,
\begin{align}
B_\pm(x,\mu) \, = \, \int\!\!d\hat s\,e^{-i y \hat s}\,B_\pm(\hat s,\mu)
\,.
\end{align}
The RG evolution of the position space jet function is described by
$B_\pm(y,\mu)=U_B(y,\mu,\mu_0)B_\pm(y,\mu_0)$, where
\begin{align}
\label{UBevolution}
U_B(y,\mu,\mu_0) \, = \,& \,
\exp\bigg[\,
\tilde\omega(\Gamma_B,\mu,\mu_0)\ln\Big(i y e^{\gamma_E} \mu_0\Big) 
\, +\, \tilde K(\Gamma_B,\gamma_B,\mu,\mu_0)
\,\bigg]\,,
\end{align}
where the functions $\tilde\omega$ and $\tilde K$ are given in
Eqs.~(\ref{omegaK}). The cusp and non-cusp anomalous dimensions read
\begin{align}
\Gamma_B^i \, = & \, \Gamma_{\rm cusp}^i\,,\quad (i=0,1,\ldots)\,,
\nn \\[1mm]
\gamma_B^0 \, = & \,4 \,C_F
\,, 
\nn \\[1mm]
\gamma_B^1 \, = & \, \, 
C_A\,C_F\,\left(\frac{1396}{27}-\frac{23}{9}\pi^2 -20\zeta_3\right) 
  + C_F\,T n_f\, \left( \frac{4}{9}\pi^2-\frac{464}{27}\right)
\,.
\end{align}
The renormalization group evolution of the hard coefficient $H_m$ is described by
$H_m(m,Q/m,\mu_m,\mu)=H_m(m,Q/m,\mu_0)U_{H_m}(Q/m,\mu_0,\mu)$ and
using the consistency condition the evolution factor $U_{H_m}$ can
be related to $U_B$ and $U_s$,
\begin{align}
U_{H_m}\Big(\frac{Q}{m},\mu_0,\mu\Big) 
\, = \, 
\Big[U_s\Big(\frac{Q}{m} y,\mu_0,\mu\Big)\Big]^{2}\,
\Big[U_B(y,\mu,\mu_0)]\Big]^{-2} \,.
\end{align}
Using the results for $U_B$ and $U_s$ in Eqs.~(\ref{UBevolution}) and 
(\ref{Usdefx}) and the relations~(\ref{omegaKrelations}) we obtain
\begin{align}
U_{H_m}\Big(\frac{Q}{m},\mu_0,\mu\Big) \, = \, 
\exp\bigg[\,
\tilde\omega(\Gamma_{c_m},\mu,\mu_0)\ln\Big(\frac{m^2}{Q^2}\Big) 
\, +\, 2\,\tilde \omega\Big(\frac{\gamma_{c_m}}{2},\mu,\mu_0\Big)
\,\bigg]
\,,
\end{align}
where the cusp and non-cusp anomalous dimensions read
\begin{align}
\Gamma_{c_m}^i \, = & \, -\,\Gamma_{\rm cusp}^i\,,\quad (i=0,1,\ldots)\,,
\nn \\[1mm]
\gamma_{c_m}^0 \, = & \,-\,4 \,C_F
\,, 
\nn \\[1mm]
\gamma_{c_m}^1 \, = & \, \,
C_A\,C_F\,\left(\!-\frac{196}{9} + \frac{4}{3}\pi^2 - 8\zeta_3\right)
+\frac{80}{9}\,C_F\,T n_f
\,.
\end{align}

\section{Fixed-Order $\tau_\alpha$ and $\rho_\alpha$ Distributions}
\label{app:distributions}

At ${\cal O}(\alpha_s^2)$ the fixed-order cumulative $\tau_\alpha$
distribution for $\mu=Q$ reads
\begin{align}
\label{taualphacumulativ}
\Sigma_{\tau_\alpha}^{\rm dijet}(\Omega) \,& = \, 
\int_0^\Omega \!\! d\tau_\alpha\,\frac{1}{\sigma_0}\,
\frac{d\sigma^{\rm dijet}}{d\tau_\alpha}
\nn\\[1mm] & 
\, = \, \theta(\Omega)\,\bigg[\,
 1 \, + \,
\left(\frac{C_F \alpha_s(Q)}{2\pi}\right)\,\Sigma_{\tau_\alpha}^{\rm dijet (1)}(\Omega)
\, + \,
\left(\frac{\alpha_s(Q)}{2\pi}\right)^2\,
\Sigma_{\tau_\alpha}^{\rm dijet (2)}(\Omega)
\, + \, \ldots
\,\bigg]
\,,
\end{align}
where [$\bar\Omega\equiv\frac{1+\alpha}{2}\Omega$]
\begin{align}
\label{taualphacumulativ2}
\Sigma_{\tau_\alpha}^{\rm dijet (1)}&(\Omega) \, = \, 
- 2 \ln^2\bar\Omega
+ \Big(-3 + 2 \ln\alpha\Big) \ln\bar\Omega
-1 + \frac{\pi^2}{3} + \frac{3}{2} \ln\alpha - \ln^2\alpha 
\,,
\nn\\[1mm] 
\Sigma_{\tau_\alpha}^{\rm dijet (2)}&(\Omega) \, = \, 
C_F^2\,\bigg[\,
2 \ln^4\bar\Omega 
+ \Big(6 - 4 \ln\alpha\Big) \ln^3\bar\Omega
+ \Big(\frac{13}{2} - 2 \pi^2 - 9 \ln\alpha 
    + 4 \ln^2\alpha\Big)\ln^2\bar\Omega
\nn\\[1mm] & \quad
+ \Big(\,\frac{9}{4} - 2 \pi^2 + 4 \zeta_3 
    + \Big(-\frac{13}{2} + 2\pi^2\Big)\ln\alpha 
    + 6\ln^2\alpha  - 2 \ln^3\alpha \Big)\ln\bar\Omega
+ 1 - \frac{3}{8}\pi^2 
\nn\\[1mm] & \quad+ \frac{5}{36} \pi^4  - 6 \zeta_3
+  \Big(-\frac{9}{8} + \pi^2 - 2\zeta_3\Big) \ln\alpha
+ \Big(\frac{17}{8} - \frac{2}{3}\pi^2\Big) \ln^2\alpha 
- \frac{3}{2}\ln^3\alpha + \frac{1}{2}\ln^4\alpha
\,\bigg]
\nn\\[1mm] &
+C_A C_F\,\bigg[\,
\frac{11}{3}\ln^3\bar\Omega
+  \Big(-\frac{169}{36} + \frac{1}{3}\pi^2 - \frac{11}{2}\ln\alpha\Big)\ln^2\bar\Omega
\nn\\[1mm] & \quad
+ \Big(-\frac{57}{4} + \Big(\frac{169}{36} - \frac{1}{3}\pi^2\Big)\ln\alpha 
  + \frac{11}{2}\ln^2\alpha + 6\zeta_3\Big)\ln\bar\Omega
+ \frac{493}{324} + \frac{85}{24}\pi^2 - \frac{73}{360}\pi^4 
\nn\\[1mm] & \quad
+ \frac{283}{18}\zeta_3 
  + \Big(\frac{57}{8} - 3 \zeta_3\Big)\ln\alpha 
+ \Big(-\frac{169}{72} + \frac{1}{6}\pi^2\Big)\ln^2\alpha 
- \frac{11}{6}\ln^3\alpha
\,\bigg]
\nn\\[1mm] &
+C_F\,T n_f\,\bigg[\!
- \frac{4}{3}\ln^3\bar\Omega
+ \Big(\frac{11}{9} + 2\ln\alpha\Big)\ln^2\bar\Omega
+  \Big(5 - \frac{11}{9}\ln\alpha - 2 \ln^2\alpha\Big) \ln\bar\Omega
\nn\\[1mm] & \quad
+ \frac{7}{81} - \frac{7}{6}\pi^2 -  \frac{22}{9}\zeta_3 
- \frac{5}{2}\ln\alpha + \frac{11}{18}\ln^2\alpha + \frac{2}{3}\ln^3\alpha 
\,\bigg]
\nn\\[1mm] &
+\frac{1}{2}\,t_{2,{\tau_\alpha}}
\,.
\end{align}

At ${\cal O}(\alpha_s^2)$ the fixed-order cumulative $\rho_\alpha$
distribution for $\mu=Q$ reads
\begin{align}
\label{rhoalphacumulativ}
\Sigma_{\rho_\alpha}^{\rm dijet}(\Omega) \,& = \, 
\int_0^\Omega \!\! d\rho_\alpha\,\frac{1}{\sigma_0}\,
\frac{d\sigma^{\rm dijet}}{d\rho_\alpha}
\nn\\[1mm] & 
\, = \, \theta(\Omega)\,\bigg[\,
1 \, + \,
\left(\frac{C_F \alpha_s(Q)}{2\pi}\right)\,\Sigma_{\rho_\alpha}^{\rm dijet (1)}(\Omega)
\, + \,
\left(\frac{\alpha_s(Q)}{2\pi}\right)^2\,
\Sigma_{\rho_\alpha}^{\rm dijet (2)}(\Omega)
\, + \, \ldots
\,\bigg]
\,,
\end{align}
where [$\bar\Omega\equiv\frac{1+\alpha}{2}\Omega$]
\begin{align}
\label{rhoalphacumulativ2}
\Sigma_{\rho_\alpha}^{\rm dijet (1)}&(\Omega) \, = \, 
-2\ln^2\bar\Omega
+ \Big(-3 + 2\ln\alpha\Big)\ln\bar\Omega
-1 + \frac{1}{3}\pi^2 + \frac{3}{2}\ln\alpha - \ln^2\alpha 
\,,
\nn\\[1mm] 
\Sigma_{\rho_\alpha}^{\rm dijet (2)}&(\Omega) \, = \, 
C_F^2\,\bigg[\,
2\ln^4\bar\Omega 
+ \Big(6 - 4\ln\alpha\Big)\ln^3\bar\Omega
+  \Big(\frac{13}{2} - \frac{4}{3}\pi^2 - 9\ln\alpha 
      + 4\ln^2\alpha\Big)\ln^2\bar\Omega
\nn\\[1mm] & \quad
+ \Big(\frac{9}{4} - \pi^2 - 4\zeta_3 + \Big(-\frac{13}{2} 
      + \frac{4}{3}\pi^2\Big)\ln\alpha + 6\ln^2\alpha 
      -  2\ln^3\alpha \Big)\ln\bar\Omega
+ 1 + \frac{3}{20}\pi^4 
\nn\\[1mm] & \quad
- 12 \zeta_3 
+ \Big(\!-\frac{9}{8} + \frac{1}{2}\pi^2 + 2\zeta_3\Big)\ln\alpha
+ \Big(\frac{17}{8} - \frac{2}{3}\pi^2\Big)\ln^2\alpha 
- \frac{3}{2}\ln^3\alpha
+\frac{1}{2} \ln^4\alpha
\,\bigg]
\nn\\[1mm] &
+C_A C_F\,\bigg[\,
+ \frac{11}{3}\ln^3\bar\Omega 
+ \Big(-\frac{169}{36} + \frac{1}{3}\pi^2 
   - \frac{11}{2}\ln\alpha \Big)\ln^2\bar\Omega 
\nn\\[1mm] & \quad
+ \Big(\!-\frac{57}{4} + 6\zeta_3
   + \Big(\frac{169}{36} - \frac{1}{3}\pi^2\Big)\ln\alpha 
   + \frac{11}{2}\ln^2\alpha \Big)\ln\bar\Omega 
+\frac{493}{324} + \frac{85}{24}\pi^2 
\nn\\[1mm] & \quad
- \frac{73}{360}\pi^4
  + \frac{283}{18}\zeta_3
+ \Big(\frac{57}{8} - 3\zeta_3\Big)\ln\alpha 
+ \Big(\!-\frac{169}{72} + \frac{1}{6}\pi^2\Big)\ln^2\alpha 
- \frac{11}{6}\ln^3\alpha
\,\bigg]
\nn\\[1mm] &
+C_F\,T n_f\,\bigg[\!
- \frac{4}{3}\ln^3\bar\Omega
+ \Big(\frac{11}{9} + 2\ln\alpha\Big)\ln^2\bar\Omega 
+ \Big(5 - \frac{11}{9}\ln\alpha - 2\ln^2\alpha\Big)\ln\bar\Omega
\nn\\[1mm] & \quad
+ \frac{7}{81} - \frac{7}{6}\pi^2 - \frac{22}{9}\zeta_3 
-\frac{5}{2}\ln\alpha  + \frac{11}{18}\ln^2\alpha + \frac{2}{3}\ln^3\alpha
\,\bigg]
\nn\\[1mm] &
+\frac{1}{2}\,t_{2,{\rho_\alpha}}
\,.
\end{align}
From the cumulative distributions in Eqs.~(\ref{taualphacumulativ}) and 
(\ref{rhoalphacumulativ}) one obtains the unintegrated distributions
$(1/\sigma_0)d\sigma^{\rm dijet}/d\tau_\alpha$ and
$(1/\sigma_0)d\sigma^{\rm dijet}/d\rho_\alpha$ by differentiation. This
entails applying the following replacement rules
[$y=\tau_\alpha$ or $\rho_\alpha$, $\kappa=\frac{1+\alpha}{2}$]:
\begin{align}
\theta(\Omega) & \,\, \to \,\, \frac{d}{d y}\Big[\theta(y)\Big] \, = \, \delta(y)
\,,
\\[1mm]
\theta(\Omega)\ln^{n+1}(\bar\Omega) & \,\, \to \,\,
 \frac{d}{d y}\Big[\theta(y)\ln^{n+1}(\kappa y)\Big]
\, = \,
(n+1)\,\kappa\,
\bigg[\,\frac{\theta(y)\ln^n(\kappa y)}{\kappa y}\,\bigg]_+
\nn\\[1mm] & \mbox{\hspace{1cm}} = \,
\ln^{n+1}(\kappa)\,\delta(y)
\, + \,
\sum_{k=0}^n \frac{(n+1)!}{(n-k)!\, k!}\,
\ln^{n-k}(\kappa)\,\bigg[\,\frac{\theta(y)\ln^ny}{y}\,\bigg]_+
\,.
\nn
\end{align}

\vskip 5mm

\section{Fourier Transform}
\label{app:fourier}

Given a momentum space variable $t$ with mass-dimension $j$ the Fourier
transform of delta-functions and plus-distributions of $t$ into position space
with the variable $y$ can be derived from the relations
\begin{align}
\int\!\! dt\,e^{-i t y} \,\theta(t)\,\frac{t^{-1+\epsilon}}{(\mu^j)^\epsilon}
\, = \, & \,\Gamma(\epsilon)\,\left(i y \mu^j\right)^{-\epsilon}
\,.
\end{align}
and
\begin{align}
\frac{\theta(t)}{t^{1-\epsilon}} \, = \, \frac{1}{\epsilon}\,\delta(t) 
\, + \,
\sum_{n=0}^\infty\,\frac{\epsilon^n}{n!}\left(\frac{\theta(t)\ln^n(t)}{t}\right)_+
\,.
\end{align}
From this result one finds
[${\cal L}^n_{t,+}=1/\mu^j[\theta(t)\ln^n(t/\mu^j)/(t/\mu^j)]_+$,
$L_y=\ln(i y e^{\gamma_E}\mu^j)$]]
\begin{align}
\int\!\! dt\,e^{-i t y}\,\delta(t) \, = \, & \,1
\,, & 
\int\!\! dt\,e^{-i t y}\,{\cal L}^2_{t,+} \, = \, & \,
-\frac{1}{3}L_y^3-\frac{\pi^2}{6}L_y-\frac{2}{3}\zeta_3
\,,
\nn\\[1mm]
\int\!\! dt\,e^{-i t y}\,{\cal L}^0_{t,+} \, = \, & \,
-\,L_y 
\,, & 
\int\!\! dt\,e^{-i t y}\,{\cal L}^3_{t,+} \, = \, & \,
\frac{1}{4}L_y^4+\frac{\pi^2}{4}L_y^2 + 2\zeta_3L_y + \frac{3}{80}\pi^4
\,,
\nn\\[1mm]
\int\!\! dt\,e^{-i t y}\,{\cal L}^1_{t,+} \, = \, & \,
\frac{1}{2}L_y + \frac{\pi^2}{12}
\,.
\end{align}
The transformation from position back to momentum space can be derived from
\begin{align}
\label{FT3}
\int\!\!\frac{dy}{2\pi}\,e^{i t y}\,\left(i y
e^{\gamma_E}\mu^j\right)^{-\epsilon}
\, = \, & \,\frac{\theta(t)}{\Gamma(\epsilon)}\,
\frac{t^{-1+\epsilon}}{(\mu^j e^{\gamma_E})^\epsilon}
\,,
\end{align}
which leads to
\begin{align}
\label{FT4}
\int\!\!\frac{dy}{2\pi}\,e^{i t y}\,\, = \, & \,\,\delta(t)
\,, &
\int\!\!\frac{dy}{2\pi}\,e^{i t y}\,L_y^3 \, = \, & 
-3{\cal L}^2_{t,+}+\frac{\pi^2}{2}{\cal L}^0_{t,+}-2\zeta_3\delta(t)
\,,
\nn\\[1mm]
\int\!\!\frac{dy}{2\pi}\,e^{i t y}\,L_y \, = \, & -{\cal L}^0_{t,+}
\,, &
\int\!\!\frac{dy}{2\pi}\,e^{i t y}\,L_y^4 \, = \, & 
\,4{\cal L}^3_{t,+}-2\pi^2{\cal L}^1_{t,+}+8\zeta_3{\cal L}^0_{t,+}
+\frac{\pi^4}{60}\delta(t)
\,,
\nn\\[1mm]
\int\!\!\frac{dy}{2\pi}\,e^{i t y}\,L_y^2 \, = \, &\, 
\,2{\cal L}^1_{t,+}-\frac{\pi^2}{6}\delta(t)
\,.
\end{align}
The cumulative distributions can be obtained directly from the position space
distributions using the relation
\begin{align}
\label{FT5}
\int\!\!\frac{dy}{2\pi}\,e^{i \Omega y}\,
\frac{\left(i y e^{\gamma_E}\kappa\right)^{\epsilon}}{i(y-i 0)}
\, = \, & \,\theta(\Omega)\,\frac{(e^{\gamma_E})^\epsilon}{\Gamma(1-\epsilon)}\,
\left(\frac{\Omega}{\kappa}\right)^{-\epsilon}
\,.
\end{align}
This leads to [$L_{y\kappa}\equiv \ln(i y e^{\gamma_E}\kappa)$, 
$L_{\Omega\kappa}\equiv\ln(\Omega/\kappa)$]
\begin{align}
\label{FT6}
\int\!\!\frac{dy}{2\pi}\,e^{i \Omega y}\,\frac{1}{i(y-i0)}\, = \, & \,\theta(\Omega)
\,, &
\int\!\!\frac{dy}{2\pi}\,e^{i \Omega y}\,\frac{L^3_{y\kappa}}{i(y-i0)} 
\, = \, & 
\theta(\Omega)\Big[
-L^3_{\Omega\kappa}+\frac{\pi^2}{2}L_{\Omega\kappa}-2\zeta_3\,\Big]
\,,
\nn\\[1mm]
\int\!\!\frac{dy}{2\pi}\,e^{i \Omega y}\,\frac{L_{y\kappa}}{i(y-i0)}  
\, = \, &  \,-\theta(\Omega)\,L_{\Omega\kappa}
\,, &
\int\!\!\frac{dy}{2\pi}\,e^{i \Omega y}\,\frac{L^4_{y\kappa}}{i(y-i0)}  
\, = \, & 
\theta(\Omega)\Big[\,
L^4_{\Omega\kappa}-\pi^2 L^2_{\Omega\kappa}+8 \zeta_3 L_{\Omega\kappa}
+\frac{\pi^4}{60}
\,\Big]
\,,
\nn\\[1mm]
\int\!\!\frac{dy}{2\pi}\,e^{i \Omega y}\,\frac{L^2_{y\kappa}}{i(y-i0)}   
\, = \, &\, 
\theta(\Omega)\Big[
L^2_{\Omega\kappa}-\frac{\pi^2}{6}
\,\Big]
\,.
\end{align}
A more general formula relevant for renormalization group improved
computations reads 
\begin{align}
\label{FT7}
\int\!\!\frac{dy}{2\pi}\,e^{i \Omega y}\,
\frac{\left(i y e^{\gamma_E}\kappa\right)^{\epsilon}}{i(y-i 0)}\,L_{y\kappa}^n
\, = \, & \,\theta(\Omega)\,
\frac{d^n}{d\epsilon^n}
\bigg[\frac{(e^{\gamma_E})^\epsilon}{\Gamma(1-\epsilon)}\,
\left(\frac{\Omega}{\kappa}\right)^{-\epsilon}\bigg]
\,.
\end{align}


\bibliography{softfctpaper}

\end{document}